\documentclass[a4paper, amsfonts, amssymb, amsmath, reprint, showkeys, nofootinbib, twoside, superscriptaddress, aps]{revtex4-1}
\usepackage{amsmath}
\usepackage{amsthm}
\usepackage[english]{babel}
\usepackage[utf8]{inputenc}
\usepackage[colorinlistoftodos, color=green!40, prependcaption]{todonotes}
\usepackage{mathtools}
\usepackage{physics}
\usepackage{xcolor}
\usepackage{graphicx}
\usepackage{subcaption}
\usepackage{adjustbox}
\usepackage{placeins}
\usepackage[T1]{fontenc}
\usepackage{bm}
\usepackage{braket}
\usepackage{multirow}
\usepackage{qcircuit}
\usepackage{hyperref}
\hypersetup{
    colorlinks=true,
    linkcolor=blue,
    urlcolor=blue,
    citecolor=blue
    }
\usepackage[capitalise]{cleveref}

\usepackage{xcolor}
\definecolor{darkblue}{rgb}{0,0,.75}
\definecolor{codegreen}{rgb}{0,0.6,0}
\definecolor{codegray}{rgb}{0.5,0.5,0.5}
\definecolor{codepurple}{rgb}{0.58,0,0.82}
\definecolor{backcolour}{rgb}{0.95,0.95,0.92}
\usepackage{listings}
\lstdefinestyle{abstyle}{
    backgroundcolor=\color{backcolour},   
    commentstyle=\color{codegreen},
    keywordstyle=\color{magenta},
    numberstyle=\tiny\color{codegray},
    stringstyle=\color{codepurple},
    basicstyle=\ttfamily\footnotesize,
    breakatwhitespace=false,         
    breaklines=true,                 
    captionpos=t,
    language= python,
    keepspaces=true,                 
    numbers=left,                    
    numbersep=5pt,                  
    showspaces=false,                
    showstringspaces=false,
    showtabs=false,                  
    tabsize=2,
}

\usepackage{tabularx}
\begin{document}

\title{Hardware-Assisted Parameterized Circuit Execution}

\author{Abhi D.~Rajagopala}
    \affiliation{Applied Math and Computational Research Division, Lawrence Berkeley National Lab, Berkeley, CA 94720, USA}
\author{Akel Hashim}
    \affiliation{Applied Math and Computational Research Division, Lawrence Berkeley National Lab, Berkeley, CA 94720, USA}
    \affiliation{Quantum Nanoelectronics Laboratory, Department of Physics, University of California at Berkeley, Berkeley, CA 94720, USA}
\author{Neelay Fruitwala}
    \affiliation{Accelerator Technology and Applied Physics Division, Lawrence Berkeley National Lab, Berkeley, CA 94720, USA}
\author{Gang Huang}
    \affiliation{Accelerator Technology and Applied Physics Division, Lawrence Berkeley National Lab, Berkeley, CA 94720, USA}
\author{Yilun Xu}
    \affiliation{Accelerator Technology and Applied Physics Division, Lawrence Berkeley National Lab, Berkeley, CA 94720, USA}
\author{Jordan Hines}
    \affiliation{Department of Physics, University of California at Berkeley, Berkeley, CA 94720, USA}
\author{Irfan Siddiqi}
    \affiliation{Quantum Nanoelectronics Laboratory, Department of Physics, University of California at Berkeley, Berkeley, CA 94720, USA}
    \affiliation{Applied Math and Computational Research Division, Lawrence Berkeley National Lab, Berkeley, CA 94720, USA}
    \affiliation{Materials Sciences Division, Lawrence Berkeley National Lab, Berkeley, CA 94720, USA}
\author{Katherine Klymko}
    \affiliation{National Energy Research Scientific Computing Center, Lawrence Berkeley National Lab, Berkeley, CA 94720, USA}
\author{Kasra Nowrouzi}
    \affiliation{Applied Math and Computational Research Division, Lawrence Berkeley National Lab, Berkeley, CA 94720, USA}

\date{\today} 

\begin{abstract}
Standard compilers for quantum circuits decompose arbitrary single-qubit gates into a sequence of physical $X_{\pi/2}$ pulses and \emph{virtual}-$Z$ phase gates. Consequently, many circuit classes implement different logical operations but have an equivalent structure of physical pulses that only differ by changes in virtual phases. When many structurally-equivalent circuits need to be measured, generating sequences for each circuit is unnecessary and cumbersome; since compiling and loading sequences onto classical control hardware is a primary bottleneck in quantum circuit execution. In this work, we develop a hardware-assisted design for executing parameterized circuits integrated with an open-source FPGA-based control system, \texttt{QubiC}. This design leverages a hardware-software co-design where software creates a circuit template, \emph{unique circuit}, by identifying structurally-equivalent circuits and \emph{peels} the relevant parameters. The hardware architecture recreates the circuit by performing real-time \emph{stitching} of the parameters to the \emph{unique} circuit on FPGA. We demonstrate the effectiveness of this work through multiple quantum characterization, verification, and validation (QCVV) protocols, including randomized benchmarking and gate set tomography, which shows a significant speedup of $17.42\times$ to $995.05\times$ in compilation time,  $64\%$ to $85\%$ of reduction in classical time with speedups of $2.81\times$ to $6.86\times$.
\end{abstract}


\maketitle

\section{Introduction}\label{sec:intro}
A quantum processing unit (QPU) can be characterized by its topology (i.e. qubit-to-qubit connectivity) and native gate set. A gate set consists of a set of native state preparations (usually chosen to be $\ket{0}$ for all qubits), a set of native single-qubit gates for all qubits, a set of native two-qubit gates between all pair of qubits, and a set of native positive operator-valued measures (POVMs; native measurements usually in the computational basis). Quantum compilers convert each quantum circuit (using a set of ordered instructions) according to the QPU's topology and gate set. Most quantum circuits have a restricted set of multi-qubit gates (e.g., using only CX or CZ gates) and arbitrary single-qubit gates. To implement arbitrary rotations in \texttt{SU(2)}, quantum compilers typically decompose single-qubit gates in terms of discrete \emph{physical} $X_{\pi/2}$ pulses (implemented via resonant Rabi-driven pulses) and arbitrary \emph{virtual}-$Z$ phase gates \cite{mckay2017efficient}. Quantum compilers achieve these decompositions by changing the phase of the subsequent physical pulse. Thus, arbitrary single-qubit gates are often $U_3$ gates and decomposed as unitaries that are \emph{parameterized} by three distinct phases:
\begin{equation}\label{eq:zxzxz}
    U_3(\phi, \theta, \lambda) = Z_{\phi - \pi/2} X_{\pi/2} Z_{\pi - \theta} X_{\pi/2} Z_{\lambda - \pi/2} ~.
\end{equation}
This $ZXZXZ$-decomposition reduces the time needed to implement arbitrary single-qubit gates since it only requires a single $X_{\pi/2}$ calibration per qubit but comes at the cost of requiring two physical pulses per single-qubit gate (for most gates). Importantly, in this manner, \emph{all} single-qubit gates are parameterized by three phases, such that implementing different single qubit gates equates to changing the phases of the physical pulses without needing to change any of the physical parameters of the $X_{\pi/2}$ pulses themselves (e.g., amplitude, frequency, pulse envelope, etc.).

Many classes of quantum circuits have \emph{structural equivalency}, i.e. they contain the same structure of physical pulses, but the virtual phases in the single-qubit gates might differ. The equivalent circuit structure can implement the same or different logical operations. In the case of many variational quantum algorithms \cite{cerezo2021variational}, such as the variational quantum eigensolver \cite{peruzzo2014variational}, the quantum approximate optimization algorithm \cite{farhi2014quantum}, and the circuits used for quantum machine learning \cite{biamonte2017quantum, benedetti2019parameterized}, the equivalent structure implements different logical operations. On the other hand, some structurally-equivalent circuits are \emph{logically-equivalent}, as is the case for noise tailoring methods such as randomized compiling (RC) \cite{wallman2016noise, hashim2021randomized}, Pauli frame randomization \cite{knill2004fault, kern2005quantum, ware2021experimental}, and equivalent circuit averaging \cite{hashim2022optimized}. Similarly, many circuits used for quantum characterization, verification, and validation (QCVV) are structurally-equivalent for a given circuit depth by design. For example, state tomography, quantum process tomography \cite{chuang1997prescription}, gate set tomography (GST) \cite{blume2013robust, blume2017demonstration}, randomized benchmarking (RB) \cite{emerson2005scalable}, cycle benchmarking (CB) \cite{erhard2019characterizing}, and many others. 

Executing quantum protocols and algorithms requires generating, compiling, and measuring a large ($\sim$100 -- 10000) number of circuits. The naive strategy is to compile each circuit independently, load them onto the control hardware, and measure them on QPU. The compilation and loading process creates a massive classical bottleneck. In many cases, this is unnecessary, since the waveform for a representative structurally-equivalent circuit is already loaded onto the hardware. For example, running a single qubit Randomized Benchmarking on eight qubits with a passive delay (for decay) of 500 $\mu$s takes about 31 seconds, of which more than 26 seconds (84\%) is from classical computation. This classical overhead becomes even more pronounced when mid-circuit measurement techniques, such as active resets, are used to minimize passive delays. This classical overhead scales linearly with the number of qubits and circuit complexity, posing a significant obstacle to quantum computation performance. To address this challenge, we introduce a hardware-software co-design technique that addresses the classical-bound bottlenecks by performing hardware-assisted parameterized circuit execution. The designs --- \emph{Read-Identify-Peel (RIP)} and \emph{Stitch} with \emph{Deft scheduler} --- which universally applies to any set of structurally-equivalent quantum circuits to improve their execution efficiency. Our approach identifies circuits with equivalent gate structures, selectively compiles \emph{unique circuits} (one from each batch of structurally-equivalent circuits), peels and stores the variable parameters for each circuit, and re-attaches at runtime while executing on the control hardware. This technique reduces the compilation time by a constant factor and adds minimal runtime overhead on resource-constrained control hardware. We demonstrate significant temporal gains for the overall runtime for several commonly used protocols, including RC, RB, CB, and GST.

\subsection{Related Work}
There are different ways to represent a quantum algorithm; however, the most widely used is using the universal quantum gate set~\cite{Nielsen_Chuang_2010} consisting of parametrized and non-parametrized quantum gates. Previous efforts have explored parameterized circuit execution primarily through software-based approaches since the development of quantum compilers such as Quipper~\cite{quipper2013green}, allowing parameterization in terms of phase, frequency, and amplitude as part of the quantum compiler design. Increased interest in variational quantum algorithms has made parameter optimization and execution essential since variational algorithms maintain common circuit structures with varying parameters iteratively computed by the optimizers. To support this, commercial mainstream quantum software such as PennyLane~\cite{pennylane2022bergholm} and IBM Qiskit~\cite{qcqiskit2024abhari} have added parameterized execution features in their software environment. In these environments, the user can specify the circuit structure and variational parameters, and the compiler and runtime system manages the binding in real-time. Other similar efforts include a partial compilation strategy by Gokhale \texttt{et al.}~\cite{partialcompilation2019gokhale}, demonstrating a $1.5\times-3\times$ speedup on variational algorithms using GRadient Pulse Engineering (GRAPE), and a one-time compilation strategy by Dalvi \texttt{et al.}~\cite{dalvi2024onetimecompilation}, demonstrating a potential speedup of up to $2.7\times, 7.7\times,$ and $2.6\times$ for representative VQE, optimal calibration, and randomized benchmarking (RB) programs, respectively. These prior works focus primarily on software-based solutions tailored to a particular type of quantum algorithm. While there are arguments for control hardware implementation of these approaches~\cite{Moll2018}, there are no such realizations. 

\subsection{Contributions}
The hardware-assisted parameterized circuit execution (PCE) design presented here takes a holistic view of circuit parameterization without restricting it to a particular class of algorithms such as VQE and differs from the above research as follows:
\begin{itemize}
	\item The front-end software, \emph{Read-Identify-Peel (RIP)}, processes not just a single circuit as in VQE but a batch of circuits and supports the automatic detection of structural equivalency in circuits in any order and peels parameters from the circuits using a single Python method. This automation is user-friendly and especially useful for QCVV algorithms, which consist of tens of thousands of circuits and hundreds of thousands of parameters.
	\item The \emph{Deft scheduler} can identify and run both in-order and out-of-order structurally-equivalent circuits. Out-of-order execution is essential for tomography applications where structural similarities are in random order in a batch of circuits.
	\item The parameter and circuit binding occur in the FPGA hardware, overlapped by the instruction pipeline and is self-contained with minimal assistance from the slower software layers during the circuit execution. Our hardware-assisted approach offers a significant advantage over previous purely software-based methods by enabling deterministic parameter binding at nanosecond latencies, thereby minimizing the  overhead.
	\item Our work also includes a profiling infrastructure. The profiling infrastructure provides the time spent on various stages during quantum algorithm execution. This highlights the bottlenecks and furnishes insight into the algorithm execution at a fine granularity instead of mere speedup using the wall clock time. Because this can be a valuable tool for understanding the quantum algorithm performance, this infrastructure is made open-source. 
\end{itemize}

Section~\ref{sec:parameter} details the \emph{RIP} software, \emph{Stitch} module hardware architecture, and the \emph{Deft scheduler}. Section~\ref{sec:time} describes the timing profile infrastructure used in profiling. Section~\ref{sec:validation} validates the hardware-assisted PCE with the experiments of different QCVV algorithms. The results demonstrates the speedup in compilation time by $17.42\times$ to $995.05\times$ (close to the theoretical limit), classical processing time by $64\%$ to $85\%$, and classical speedups of $2.81\times$ to $6.86\times$ for different QCVV algorithms.

\section{Parameterized Circuit Execution}\label{sec:parameter} 

\begin{figure*}
    \centering
    \includegraphics[width=.95\textwidth]{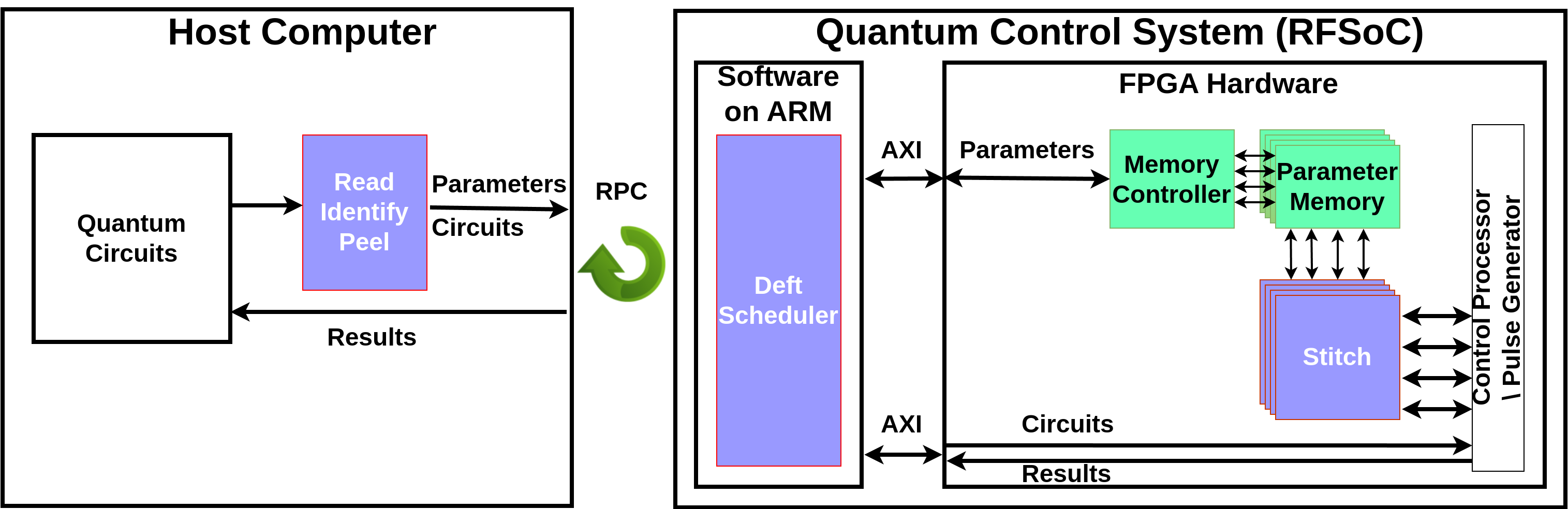}
    \caption{Block Diagram showing different parts of hardware-assisted parameterized circuit execution (PCE) with \emph{Read-Identify-Peel (RIP)}
     software on the host computer, \emph{deft scheduler} on the control software, and \emph{stitch module} on the FPGA. The \emph{RIP} on host computer separates the circuits from its parameters and compiles \emph{unique circuits} and transfers them to the control system. The \emph{Deft scheduler} running as a software on control system schedules circuits and the appropriate parameters. The parameters and circuits are stitched back together in control system FPGA in real time. The results are transferred back to the host computer at the end of circuit execution.}
    \label{fig:paramArch}
\end{figure*}
Improving classical efficiency during quantum circuit execution motivates this work. To reduce classical bottlenecks, we introduce the hardware-assisted parameterized circuit execution (PCE) design, which exploits structural equivalency in a batch of circuits to improve efficiency. As shown in Fig.~\ref{fig:paramArch}, this hardware-software co-design utilizes two interconnected computational systems: a general-purpose computer (Host Computer) and a radio frequency system-on-a-chip (RFSoC). The RFSoC is an AMD Zynq UltraScale+ RFSoC ZCU216 evaluation kit~\cite{zcu216} with an embedded ARM processor and an FPGA. On the host computer, Python-based \emph{Read-Identify-Peel (RIP)} software identifies structurally-equivalent circuits, extracts parameters from single-qubit gates, and compiles \emph{unique circuits}. The \emph{Deft scheduler} software on the ARM processor of RFSoC orchestrates circuit execution, while the \emph{Stitch} module on the FPGA performs parameterized execution. The PCE design integrates with an open-source quantum control system, \texttt{QubiC}~\cite{xu2023qubic}, for full-stack support. \texttt{QubiC} is an FPGA-based distributed architecture with multiple cores to generate arbitrary waveforms for control and read measurements on superconducting quantum devices. The software layer implements an intermediate representation (\texttt{QubiC-IR}) and a modular compiler toolchain for quantum circuit execution. The PCE design integrates with \texttt{QubiC} at multiple layers: the \emph{Read-Identify-Peel (RIP)} software sits atop the \texttt{QubiC} compiler stack, the \emph{Deft scheduler} interfaces with the control system runtime software, and the \emph{Stitch} module interacts with the distributed processor. The following sections detail the design and functionality of the \emph{RIP} software, the \emph{Deft scheduler}, and the \emph{Stitch} hardware architecture.

\subsection{\emph{Read-Identify-Peel (RIP)}}
\label{subsec:rip}

\begin{figure*}
    \centering
    \includegraphics[width=.750\textwidth]{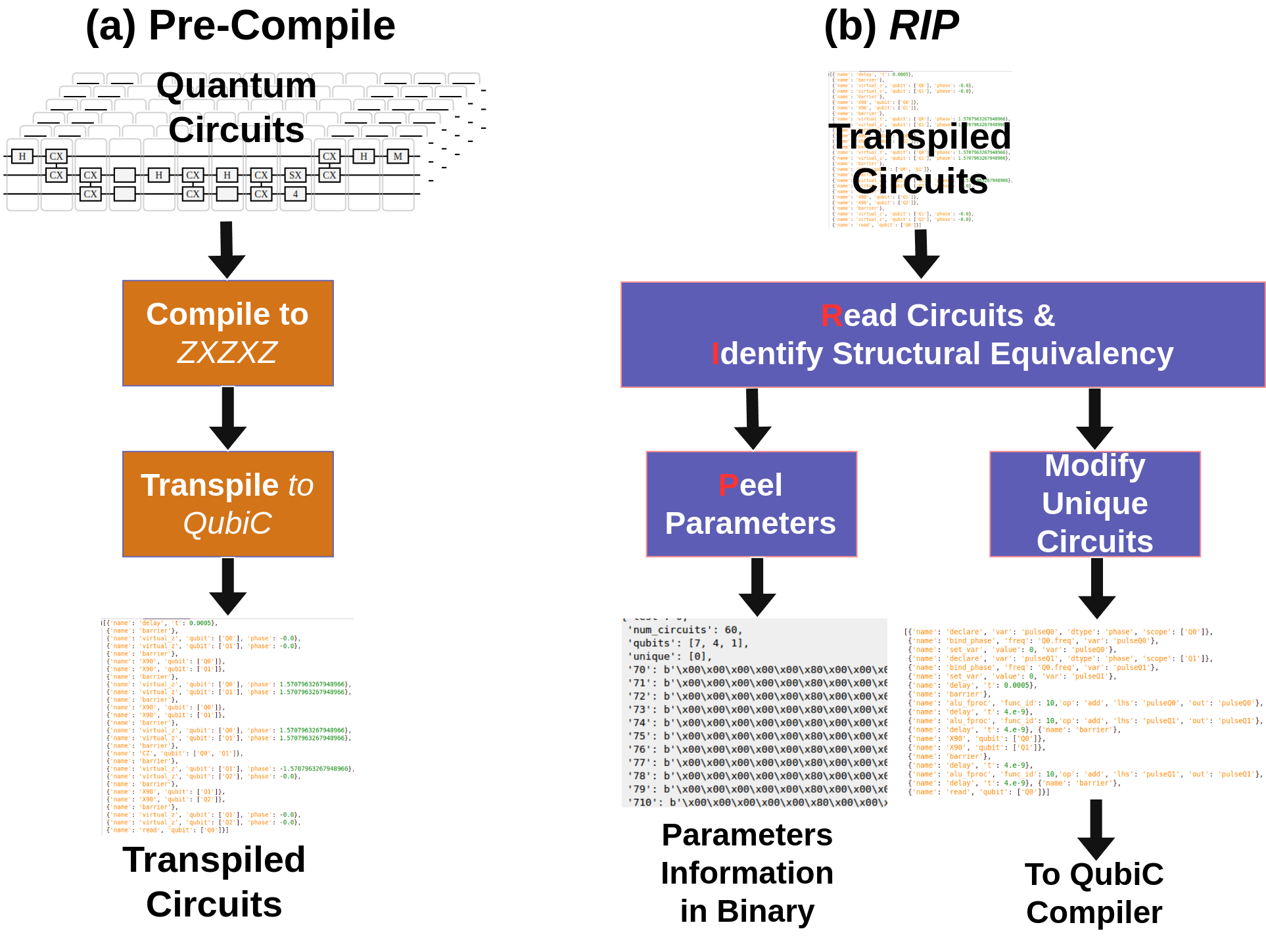}
    \caption{\textbf{Stages of \emph{Read-Identify-Peel (RIP)} software process}. 
    \textbf{(a)} The transformation steps of quantum circuits written in  \texttt{openQASM3}, \texttt{Qiskit}, \texttt{True-Q}, or \texttt{pyGSTi} into \texttt{QubiC} native gates. The circuits are compiled using Equation~\ref{eq:zxzxz} and transpiled into \texttt{QubiC} native gate format. \textbf{(b)} The parameterization steps of identifying structural equivalency in a batch of transpiled circuits and creating \emph{unique circuits} (one from each structurally-equivalent batch), peeling the phase information from the circuits, and modifying the \emph{unique circuits} to request the peeled phases in real-time on the FPGA.}
    \label{fig:ripstage}
\end{figure*}

As a crucial first step in the hardware-assisted PCE design, the \emph{Read-Identify-Peel (RIP)} process, implemented in Python, runs on the Host Computer and prepares circuits for hardware-assisted parameterized circuit execution. As depicted in Figure~\ref{fig:ripstage}, the overall process consists of two process: a) \emph{Pre-compile} b) \emph{RIP}. The \emph{Pre-compile} process prepares the circuit for \emph{RIP}, which analyzes \emph{pre-compiled} circuits, identifies structural equivalencies, extracts parameters, and modifies the circuits for efficient execution on the control system. Designed to be generic and adaptable, the process supports circuits from various quantum programming languages, including  \texttt{openQASM3}~\cite{Cross_2022}, \texttt{Qiskit}~\cite{qiskit2024}, \texttt{True-Q}~\cite{trueq}, and \texttt{pyGSTi}~\cite{osti_1543289}, and integrates with the \texttt{QubiC} control system's \texttt{QubiC-IR}~\cite{qubicir} intermediate representation. The \emph{pre-compile} stage compiles circuits from different quantum programming languages into single and two-bits gates, where the single-qubit gates are compiled according to Eq.~\ref{eq:zxzxz}. A transpiler converts these compiled gates into \texttt{QubiC} native gate abstraction which serves as the input to the \emph{RIP} process.

The Read-Identify method analyzes the structural equivalency within a transpiled batch of circuits. It processes each circuit by constructing a directed acyclic graph (DAG) to represent the sequence of gates applied to each qubit. Each node in the graph corresponds to a gate operation, and edges indicate the temporal order of the gate execution. By comparing these graphs, the Read-Identify method identifies circuits with identical structures. The computational complexity of this equivalency analysis is $\mathcal{O}(n^2)$, where $n$ is the number of circuits. A circuit is considered \emph{unique}, if its graph structure and metadata (except parameter values such as virtual phases) is distinct from all previously analyzed circuits. Otherwise, the circuit is identified as structurally-equivalent with a previously identified \emph{unique circuit}. Thus, a \emph{unique circuit} serves as a template for all circuits with the same structure. For each \emph{unique circuit}, a new sublist is created to store its index and the indices of any structurally-equivalent circuits encountered later. After processing the entire circuit batch, a flattened list of sublists, which contains the \emph{unique circuits} and its corresponding structural equivalent circuits provides the circuit execution order for the \emph{Deft scheduler}. By default, the Read-Identify method can effectively identify structural equivalencies in circuits in any order, not just adjacent circuits, leading to \emph{out-of-order} circuit execution. However, the user can force an in-order equivalence to preserve the input circuit batch order depending on the algorithm.

The Peel method extracts relevant parameters from the entire batch of circuits, both \emph{unique} and structurally-equivalent circuits. Using the graphs from the Read-Identify method, Peel process extracts the \emph{virtual}-$Z$ phases from the single-qubit gates, which can vary across different circuits within the batch. The extracted parameters are organized into a dictionary value, where keys are a combination of the target qubit and circuit index from the Read-Identify method sublist. This dictionary is binarized to optimize data transfer between the Host Computer and the control system. This binarization technique provides a significant performance improvement — up to a $15\times$ speedup — compared to transmitting a non-binarized dictionary.

The final stage of the \emph{RIP} process involves modifying the \emph{unique circuits}' instructions to enable real-time parameter stitching. This modification process replaces each \emph{virtual}-$Z$ instruction, from which parameters have been extracted, with an \texttt{alu\_fproc} instruction from the \texttt{QubiC} instruction set, that requests the corresponding parameter value from the \emph{Stitch} module in real-time on the FPGA. The \texttt{alu\_fproc} instruction is defined as follows:

\begin{lstlisting}[label={lst:qubifroc},caption={\texttt{QubiC} function processor instruction through which distributed processor requests data from function processor (\emph{Stitch logic}) and stores it in a variable.}, style={abstyle}, numbers=none]
    'name': 'alu_fproc', 'func_id':10, 'op':'add', 'lhs':varname, 'out':varname
\end{lstlisting}
Here, \texttt{func\_id} of $10$ is the identifier by which \emph{Stitch} module recognizes it as a parameter requests from the \texttt{QubiC} distributed processor, and \texttt{varname} is a variable used to hold the parameter value in the distributed processor's core (one per physical qubits). These instructions, specific to the \texttt{QubiC} distributed processor, enable the parameterized circuits to efficiently access the correct parameter values from the \emph{Stitch} module on the FPGA during circuit execution. The entire modified instruction is shown in Listing~\ref{lst:modc} in Appendix~\ref{subsec:stitchcode}, and Listings~\ref{lst:origcirc} and~\ref{lst:modicirc} in Appendix~\ref{subsec:circmod} shows circuits with structural similarities and the modified \emph{unique} circuit. In the \texttt{QubiC} architecture framework, the \emph{Stitch} module acts as a centralized Function Processor (\texttt{fproc})~\cite{fruitwala2024distributed} with parallel data bus, providing a mechanism for multiple distributed processors to access circuit parameters in real-time. This real-time access is essential for minimizing latency and ensuring hardware-assisted efficient execution of the parameterized circuits. More details about the \emph{Stitch} module is provided in Section~\ref{subsec:stitch}.

The \emph{RIP} process integrates seamlessly with the \texttt{QubiC} compiler stack, slotting between the \emph{pre-compile} and \texttt{QubiC} compilation stages. Specifically, \emph{RIP} takes the output of the \emph{pre-compile} stage, which involves translating quantum programs into a \texttt{QubiC-IR} intermediate representation using Equation~\ref{eq:zxzxz}, and its output feeds into the subsequent \texttt{QubiC} compilation stages. To support quantum applications utilizing this alternative compilation flow, we introduce a new Python API method, \texttt{pce}, and modify an existing API method in the \texttt{QubiC} software stack. The \texttt{pce} method, as shown in Listing~\ref{lst:pce} in Appendix~\ref{subsec:ripcode}, takes the pre-compiled circuit, the qubit map (describing the physical topology of the quantum processing unit), and the channel map (containing information about the distributed processor cores and physical interfaces) as input arguments. It outputs a binarized parameter dictionary and a list of modified \emph{unique circuits}. To interface this new functionality into \texttt{QubiC} control software, the \texttt{build\_and\_run\_circuits} method in the \texttt{QubiC} API is modified to \texttt{build\_and\_run\_paracircuits} as shown in Listing~\ref{lst:buildrun} in Appendix~\ref{subsec:runcode}. This modification supports transmitting parameter dictionary and circuit list as input to the \emph{Deft scheduler}, running on the RFSoC, using remote procedure call (RPC).

\subsection{\emph{Stitch}}
\label{subsec:stitch}
The \emph{Stitch} process, an essential component of the hardware-assisted PCE architecture, re-attaches the parameters extracted by the \emph{RIP} process on the FPGA, enabling real-time parameterization during circuit execution by storing the parameters in a dedicated memory and providing them to the \texttt{QubiC} distributed processor as needed. The \emph{Stitch} module design comprises three subcomponents: the memory controller, parameter memory, and stitch logic. The memory controller enables the read/write of parameters from the ARM processor, the parameter memory stores the parameter values, and the stitch logic interacts with the \texttt{QubiC} distributed processor to provide parameters when required. The \emph{Stitch} module is designed for modularity and extensibility and can be readily adapted for systems with varying numbers of qubits, promoting flexibility and scalability in the PCE architecture. The current design, implemented in a combination of Verilog and System Verilog~\cite{sysVerilog}, operates at a clock speed of 500 MHz and has been tested for up to eight physical qubits.

The \emph{Stitch} module resides on the control system's hardware platform, a Zynq UltraScale+ RF-SoC ZCU216 evaluation kit. This system-on-a-chip platform features ARM processors or Processing System (PS) and an FPGA or Programmable Logic (PL) operating in different frequency domains to optimize performance and power consumption~\cite{mpsoc2019book}. The \emph{Stitch} module employs a memory controller that facilitates communication between the ARM processor, running at 1.33 GHz, and the parameter memory on the FPGA, running at 500 MHz. This communication occurs via the Full Power Domain AXI bus operating at 100 MHz~\cite{XilinxUG1085}. The memory controller has two subcomponents: an AXI to local bus converter and a memory switch. The AXI to local bus converter translates the 32-bit AXI bus to a 14-bit local bus, addressing the difference in frequency and address widths between the AXI bus and the parameter memory. The memory switch then decodes the local bus address to access the appropriate parameter memory. This decoding process utilizes the first 3 MSBs of the 14-bit address to select one of the eight parallel memories and the remaining 11 bits to address a specific location within that memory, providing access to $2048$ memory locations per memory. The memory switch also manages the write enable signal for the parameter memories and read valid signal for the AXI to local bus converter.

The parameter memory holds the circuit parameters and should support multiple kilobytes of memory for each qubit. The design choice in the current implementation is the Block RAM (BRAM)~\cite{XilinxUG573}, due to its 32-kilobit block size, which are modular and better to maintain higher operating frequency with lower timing issues during routing. Alternatively, an Ultra RAM~\cite{XilinxUG573}, with a 288-kilobit block, will be underutilized and difficult to route, and a distributed RAM using memory LUT (SLICEM)~\cite{XilinxUG574} would require significant resources. The parameter memory comprises eight parallel BRAMs, one for each physical qubit in the current implementation. Each BRAM has a capacity of 8 kilobytes and can store up to 2048 32-bit parameter values per circuit. Each memory is a true dual-port (TDP) design~\cite{XilinxUG573}, with one interface dedicated to the memory controller for writing parameters and the other to the stitch logic for reading parameters during circuit execution. This dual-port configuration enables efficient concurrent access for both writing and reading parameters. Importantly, the \emph{Deft scheduler} avoids memory race conditions with mutually exclusive memory write and circuit execution operations. During parameterized circuit execution, the memory controller writes the circuit parameters to the individual BRAM, and the stitch logic reads from the memory based on requests from the distributed processor. The remaining read and write capabilities of the TDP design, such as reading from the memory controller side or writing from the stitch logic side, are reserved for debugging purposes, providing valuable flexibility for testing and troubleshooting.

The stitch logic block is responsible for the hardware-assisted parameterization process, interacting with the parameter memories and the \texttt{QubiC} control system's distributed processor. The design is modular and the current design integrates with eight physical qubits, featuring eight parallel interfaces for the parameter memories and eight \texttt{fproc} interfaces for the distributed processor, which requests a parameter via the \texttt{fproc} interface using an 8-bit ID. The stitch logic fetches the corresponding parameter value from the appropriate parameter memory and returns it through the \texttt{fproc}'s 32-bit data interface. To further optimize performance, the stitch logic employs a prefetch mechanism that anticipates future parameter requests and fetches them in advance, reducing latency and achieving a parameter request completion within two clock cycles ($4$ ns). Since quantum circuits are typically executed multiple times (shots) with the same set of parameters, the stitch logic tracks the number of parameters in each circuit and automatically repeats the parameter delivery for the specified number of shots. Moreover, the stitch logic can repeat a partial set of parameters or switch to a different set based on control codes provided by the \emph{Deft scheduler} before circuit execution. This flexibility enables dynamic parameter adjustments during circuit execution. The stitch logic also preserves the existing mid-circuit measurement and feed-forward functionality of the \texttt{QubiC} system, which allows for conditional execution of gates based on measurement outcomes. It distinguishes between parameter requests and mid-circuit measurement requests from the control processor using the \texttt{core\_id} parameter in the \texttt{fproc} interface. For instance, \texttt{core\_id} values between 0 and 7 indicate mid-circuit measurement requests for the corresponding qubits, while the value of 10 indicates a parameter request. These \texttt{core\_id} values are encoded in the circuit during the circuit modification stage in the \emph{RIP} process.

The design of the stitch logic focused on two main objectives: minimizing FPGA resource utilization and achieving a 500 MHz operating frequency. This frequency requirement stems from the \texttt{QubiC} control system, whose distributed processor operates at 500 MHz and with which the stitch logic interfaces. Resource utilization impacts the availability of resources for other logic elements, the maximum achievable operating frequency, and routing congestion on the FPGA. With these design constraints, the stitch logic reuses existing IPs from the \texttt{QubiC} gateware, such as the clock converter IP~\cite{XilinxDS768}, and implements a streamlined design. The utilization report, generated from Vivado~\cite{XilinxUG906}, for a 8-qubit system is shown in Table~\ref{tab:stitchutil}. The overall resource utilization is as follows: 1695 Look-Up Tables (LUTs), 2236 Registers, and 16 blocks of 32k BRAM. This utilization corresponds to $0.4\%$ of the available LUTs, $0.26\%$ of the available registers, and $1.48\%$ of the available BRAMs. Block RAMs are a critical resource on the FPGA, with a total of $1080$ blocks available in the ZCU216 development board. The current design's efficient utilization of $2$ BRAM blocks per qubit ensures sufficient resources for other control system components, which use extensive memory resources for pulse generation. By achieving these low resource utilization figures, the \emph{Stitch} module successfully meets its design objectives and operates at the target $500$ MHz frequency.

\begin{table*}[!t]
    \centering
    \caption{Resource utilization of the \emph{Stitch} module for an 8-qubit system, broken down by component: bus converter, memory switch, parameter memory for all 8 qubits, and stitch logic.}
    \label{tab:stitchutil}
\begin{tabularx}{1.0\textwidth} { 
      >{\raggedright\arraybackslash}X 
      || >{\raggedleft\arraybackslash}X 
      | >{\raggedleft\arraybackslash}X 
      | >{\raggedleft\arraybackslash}X 
}
      \textbf{Module}&\textbf{LUTs}&\textbf{Registers}&\textbf{Block RAMs}\\
    \hline
    \hline
Total Available Resources&	425820 (100\%)&	850560(100\%)&	1080(100\%)\\
\hline
AXI to local bus&	154 (0.04\%)&	231 (0.03\%)&	0 (0.00\%)\\
Memory Switch&	77 (0.02\%)&	4 (0.00\%)&	0 (0.00\%)\\
Parameter Memory&	160 (0.04\%)&	0 (0.00\%)&	16 (1.48\%)\\
stitch Logic&	1304 (0.31\%)&	2001 (0.24\%)&	0 (0.00\%)\\
\hline
\emph{Stitch} module Total&	1695 (0.40\%)&	2236 (0.26\%)&	16 (1.48\%)\\
    \end{tabularx}
\end{table*}

\subsection{\emph{Deft scheduler}}

The \emph{Deft scheduler} orchestrates the execution of parameterized circuits by coordinating the interaction between the \emph{RIP} process, the \emph{Stitch} module, and the \texttt{QubiC} control system. Implemented in Python, the scheduler runs on the ARM processor of RFSoC control hardware. It integrates seamlessly with the existing \texttt{QubiC} runtime software with new methods for managing parameterized circuit execution. These additions allow users to leverage the PCE functionality without disrupting the existing \texttt{QubiC} control software. On the control system side, the scheduler introduces low-level memory drivers for loading the parameter values. It uses the IP dictionary from PYNQ~\cite{pynq} to determine the memory map address of the \texttt{AXI to the local bus converter} on the \emph{Stitch} module and writes the parameter value to the appropriate parameter memory using the address offset. It also reuses existing \texttt{QubiC} methods for tasks such as loading the pulse and frequency envelopes, which define the shape and timing of the control signals applied to the qubits, loading circuit instructions, triggering circuit runs, and collecting measurement data.

During quantum circuit execution, the \emph{Deft scheduler} receives binarized parameter dictionary and the compiled \emph{unique circuits} from the host computer via a remote procedure call (RPC). The scheduler de-binarizes the parameter dictionary to extract the circuit order list, the \emph{unique circuit} list, and the corresponding parameter values. The scheduler uses the circuit order to fetch the circuit index and loads the parameters, one physical qubit at a time, into the designated parameter memory on the PL. If the circuit index is also in the \emph{unique} circuit lists, then it loads the corresponding circuit template. This selective loading minimizes redundant loading of structurally-equivalent circuits and reduces the classical execution time. Once the parameters and circuits are loaded, the scheduler proceeds with the standard \texttt{QubiC} software operations, including loading user-defined settings, executing the circuits, retrieving measurement data from the PL, and transmitting the results back to the host computer. These sequence of operations continue for each index in the circuit order list.
\section{Time Profiling}\label{sec:time}

The hardware-assisted parameterized circuit execution (PCE) design should reduce the execution time of quantum circuits by selective compilation and loading of \emph{unique} circuits. As seen in Sec.~\ref{sec:parameter}, a quantum circuit execution consists of multiple steps over different computational domains, such as the host computer and the control system. Understanding the efficacy of PCE requires detailed time profiling of each step in the execution process. Thus, we added a time profiling infrastructure to \texttt{QubiC} to measure the time at different stages.

\subsection{Time Profiling Infrastructure}
\label{subsec:timeprof}

The time profiling presented here occurs at three layers: the application, the host computer software, and the control system software. Each layer has multiple classical computation stages; in some cases, each stage has sub-stages of operations. During a circuit execution, these stages and sub-stages may run for varying iterations. For example, to run a single circuit for 100 shots, the compilation and the circuit load stages execute once, whereas the run circuit stage executes for 100 iterations (once per shot). The time profiler captures the time spent on all the stages and sub-stages at different layers for varying iterations into a Python dictionary. For ease of use, the profiling infrastructure integrates with the control software layer and allows the user to select the time profiling for all layers from the application layer using the \texttt{build\_and\_run\_paracircuits} as shown in Listing~\ref{lst:buildrun} in Appendix~\ref{subsec:runcode}.

\textbf{Application layer} is the front end of the circuit execution, runs on a Jupyter Notebook on the host computer. It is typically responsible for circuit creation, pre-compilation, parameterization, control software initiation for compile and run, and result display. The different profiling stages in this layer are as follows:

\begin{itemize}
    \item Pre-compile: This stage encompasses creating or reading circuits, compiling them into a gate-based representation, and transpiling them to the native control system gate format. This process is detailed in Section~\ref{subsec:rip} and Fig.~\ref{fig:ripstage}(a). The \emph{pre-compile} stage differentiates from the compile stage in the host computer software layer, where the latter converts the circuits into pulse definitions. The pre-compile stage has the following sub-stages:
    \begin{itemize}
        \item Get circuit: This sub-stage generates circuits for the quantum experiment or reads a pre-generated circuits from a file.
        \item Transpile: This sub-stage converts the circuit, which may be represented in various quantum programming languages, into the native control software gate format, based on Eq.~\ref{eq:zxzxz}. Note, the stages \emph{Compile} and \emph{Transpile} in Fig.~\ref{fig:ripstage}(a) is combined into this sub-stage to avoid confusion with the \emph{Compile} stage that occurs later during \texttt{QubiC} compilation.
    \end{itemize}
    \item \emph{RIP}: This stage performs software parameterization, identifying structural equivalency, extracting phases, and modifying the circuit for PCE, as described in Section~\ref{subsec:rip}.
    \item Active: This optional stage converts a circuit from a passive reset delay ($500 \mu$s in the current QPU) to an active reset circuit. The experiments in this manuscript use passive reset delay. 
    \item Build and Run: This stage transfers control from the application layer to the host computer software layer and receives the quantum measurement data from the control system. The sub-stages of this layer is discussed in the host computer software layer.
    \item Total: This represents the complete execution time from the start of circuit creation to the display of the result for an experiment measured from the Jupyter notebook.
\end{itemize}

The application layer also includes methods to initialize and control time profiling in all three layers. Users can access these methods by importing the profiling module. As shown in Listing~\ref{lst:profile}, the method \texttt{setClientProf()} activates client-side profiling on the host computer software layer, and \texttt{logjson()} logs the profile dictionary into JSON format. Server-side profiling, which activates time profiling on the control software layer, is controlled by the \texttt{server\_profile} argument to the \texttt{build\_and\_run\_paracircuit} method, as shown in Listing~\ref{lst:buildrun} in Appendix~\ref{subsec:profcode}. A value of zero indicates no profiling on the control software, while a value of 1 enables profiling.

The profiling in the \textbf{Host Computer Software Layer} is triggered by the application layer during circuit execution. This layer also runs on the host computer as a Python module. It initiates circuit execution by compiling and assembling the circuit and transferring the output to the software running on the control hardware, acting as a client in a remote procedure call (RPC). After circuit execution, it reads the data from the RPC server running on the control hardware and passes it to the application layer. This layer has the following stages and sub-stages:
\begin{itemize}
    \item Compile: This stage converts the circuit from the native control software gate format into control hardware-specific assembly instructions. This involves translating the gate-level representation of the circuit into pulse instructions and resolving the timing.
    \item Assemble: This stage converts the assembly language instructions into machine code, which is the binary representation that can be directly executed by the \texttt{QubiC} distributed processor. 
    \item RunAll on Host: This stage encompasses the time to run the circuits and obtain the result data, equivalent to \emph{Build and Run} in the application layer.
    \begin{itemize}
        \item Run on Host: This sub-stage measures the runtime of the circuits from the perspective of the host computer, including the time spent on communication with the control hardware via RPC.
        \item Data Sort: This sub-stage converts the raw IQ data received from the control hardware into a distribution of bitstrings, representing the measurement outcomes of the quantum circuit.
    \end{itemize}
    \item Client/Server: This post-processing stage calculates the time taken for data transmission between the RPC client (on the host computer) and the RPC server (on the control hardware). This measurement helps to understand the communication overhead from the overall execution time.
\end{itemize}

The \textbf{Control Software Layer} runs on the ARM processor of the control hardware platform (AMD ZCU216 RFSoC) as a Python module and receives the circuits and execution parameters from the host computer software layer via the RPC. It interacts with the FPGA to load all the circuit elements, run the circuits, and collect the data. This process has the following stages:
\begin{itemize}
    \item Load Batch: This stage transfers all the necessary parameters for circuit execution from the ARM processor to various memories in the FPGA. This includes:
    \begin{itemize}
        \item Load circuit: This sub-stage transfers the compiled and assembled quantum circuits from the ARM processor to the instruction memory in the FPGA.
        \item Load definition: This multi-stage process loads various circuit parameters, such as the signal envelopes, which define the shape and timing of the control pulses applied to the qubits and the frequencies of those pulses. Thus, it is divided into following sub-stages:
        \begin{itemize}
            \item Load envelope: This sub-stage transfers the signal envelope data to the FPGA's memory.
            \item Load frequency: This sub-stage transfers the frequency parameters to the FPGA's memory.
            \item Load zero: This sub-stage clears the command buffer in the FPGA, ensuring that any previous commands do not interfere with the current circuit execution.
        \end{itemize}
    \end{itemize}
    \item Load para: This stage transfers the parameters extracted by the \emph{RIP} process (the "peeled" parameters) to the parameter memory in the FPGA. These parameters are used by the \emph{Stitch} module in FPGA during circuit execution.
    \item Run Batch: This stage starts with the trigger to initiate circuit execution and ends with retrieving the quantum measurement data.
    \begin{itemize}
        \item Start Run: This sub-stage measures the runtime of the circuit on the FPGA, capturing the time taken for the actual execution of the quantum operations.
        \item Get data: This sub-stage measures the time to read the quantum measurement data from the FPGA, which involves transferring the measurement outcomes from the FPGA's memory to the ARM processor.
    \end{itemize}
    \item \emph{Stitch}: This post-processing measurement is the time taken by the \emph{Stitch} module to serve parameter requests from the distributed processor within the FPGA. This measurement helps to evaluate the efficiency of the hardware-assisted parameterization process.
\end{itemize}

\section{Validation, Benchmarking, and Characterization}\label{sec:validation}

This section validates the effectiveness of the hardware-assisted parameterized circuit execution (PCE) design in improving the execution efficiency on various classes of benchmarking and characterizing quantum algorithms, particularly those with a large number of structurally-equivalent circuits. A prototypical example is quantum process tomography (QPT), where circuits are identical except for changes in preparation and measurement basis. These basis rotations can be expressed as combinations of a single physical gate (e.g., $X_{\pi/2}$) and \emph{virtual}-$Z$ gates. Consequently, many QPT circuits share the same structure of physical pulses, differing only in their virtual phases. The PCE design leverages this structural similarity by compiling only the unique physical pulse sequences in the host computer and efficiently managing the virtual phases in the FPGA, leading to a substantial reduction in compilation overhead and optimized parameter handling. We demonstrate the performance advantage of the PCE design against standard software for randomized compiling (RC)~\cite{wallman2016noise, hashim2021randomized}, randomized benchmarking (RB)~\cite{emerson2005scalable, knill2004fault, magesan2011scalable}, cycle benchmarking (CB)~\cite{erhard2019characterizing}, and gate set tomography (GST)~\cite{blume2013robust, blume2017demonstration, nielsen2021gate}. The experiment setup includes a host computer, a desktop with Intel i9-11900K processor, the control system, an AMD ZCU216 RFSoC evaluation board running QubiC, and a superconducting 8 qubit QPU at DOE's Advanced Quantum Testbed (AQT).


\subsection{Randomized Compiling}
\begin{figure*}[!t]
    \centering
    \includegraphics[width=.75\textwidth]{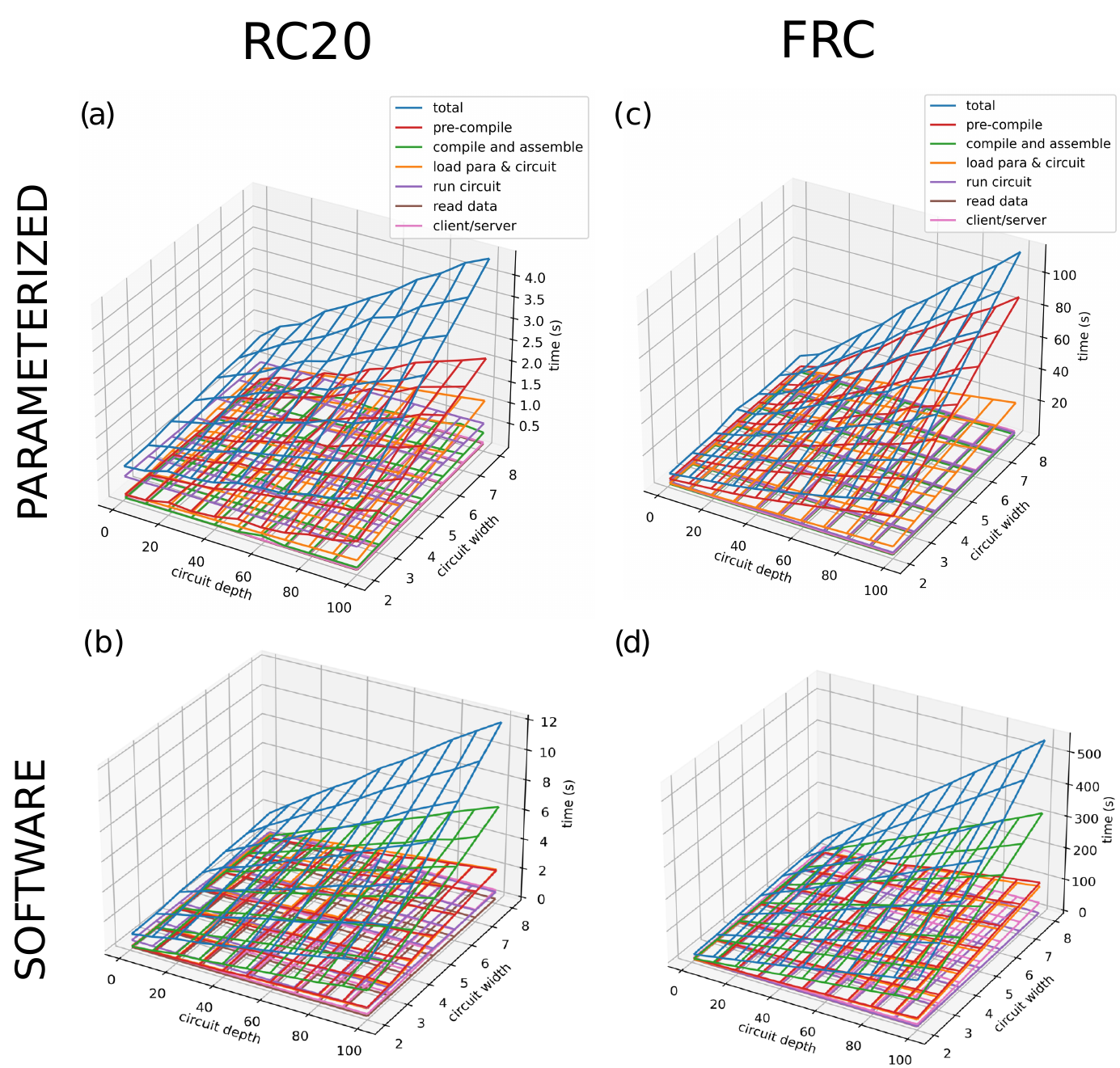}
    \caption{\textbf{Time Profiling.}
    We compare a breakdown of the execution time for performing RC on circuits of various widths (2 -- 8 qubits) and depths (1 -- 100 cycles of two-qubit gates) using the software and PCE design. For the software design, we implement RC using (a) 20 and (c) 1000 different randomizations over a total of 1000 shots. The PCE design gives a $2.81\times$ and $4.32\times$ classical time speedup (Eq.~\ref{eq:csu} over software RC for (b) $N = 20$ and (d) $N = 1000$ randomizations, respectively.
    }
    \label{fig:time}
\end{figure*}

Randomized compiling (RC) is a technique that enhances the performance and stability of quantum algorithms by converting coherent and non-Markovian errors into stochastic errors. It achieves this by randomizing the gates in a circuit using virtual twirling groups while preserving the overall ideal unitary of the circuit~\cite{wallman2016noise}.

To benchmark the runtime improvements of PCE for classes of equivalent circuits, we profile the execution time of random circuits at varying depths and widths, measured with RC, as shown in Figure~\ref{fig:time}. The circuit depths range from 1 to 100, defined by the number of two-qubit gate cycles, and the widths range from 2 to 8 qubits. These variations in depth and width allow us to assess the performance of PCE across a range of circuit complexities. We measure each circuit under two RC scenarios: one with $N=20$ randomizations (referred to as RC20) with $50$ shots per randomization and another with $N=1000$ randomizations (referred to as \emph{fully randomly compiled (FRC)} limit~\cite{fruitwala2024hardware},  with 1 shot per randomization. We generate RC circuits using \texttt{True-Q}~\cite{trueq} for both the standard software and the PCE implementations. The software implementation follows the existing \texttt{QubiC} software infrastructure to generate sequences and upload them to the control hardware for each randomization. The PCE implementation utilizes the \emph{RIP} and \emph{Stitch} design to optimize circuit execution. The detailed experimental parameters for the both the RC cases are listed in Appendices~\ref{subsec:paramrc20} and~\ref{subsec:paramfrc}, respectively.

Figures~\ref{fig:time}(a) and (b) present the profiling results for the parameterized and software RC with $N=20$ randomizations (RC20), respectively, while Figures~\ref{fig:time}(c) and (d) show the results for $N=1000$ randomizations (FRC). We observe that parameterized RC provides a significant classical speedup (defined in Eq.~\ref{eq:csu}) over software RC, achieving approximately a $2.81\times$ speedup for RC20 and a $4.32\times$ speedup for FRC. This improvement is attributed to the reduced compilation overhead and optimized parameter handling in the PCE design. Since in an RC circuits of $n$ randomization, all $n$ circuits are structurally equivalent, the PCE design recreates the entire $n$ circuits by compiling just a single \emph{unique} circuit per depth and width. As shown in Table~\ref{tab:summary}, PCE recreates $1540$ and $77,000$ circuits for RC20 and FRC, respectively, by compiling only $77$ \emph{unique} circuits and managing parameters. For software RC, the primary bottleneck is the compile and assemble stage, which involves compiling the entire batch of circuits from \texttt{QubiC-IR} to the low-level machine code. This process is computationally expensive and time-consuming, especially for increased randomizations. In contrast, for parameterized RC, the main bottleneck is the \emph{pre-compile} stage, which stems from the initial compilation of \texttt{True-Q} circuits into \texttt{QubiC} native gates.

\subsection{Randomized Benchmarking}
\begin{figure*}[!t]
    \centering
    \includegraphics[width=.90\textwidth]{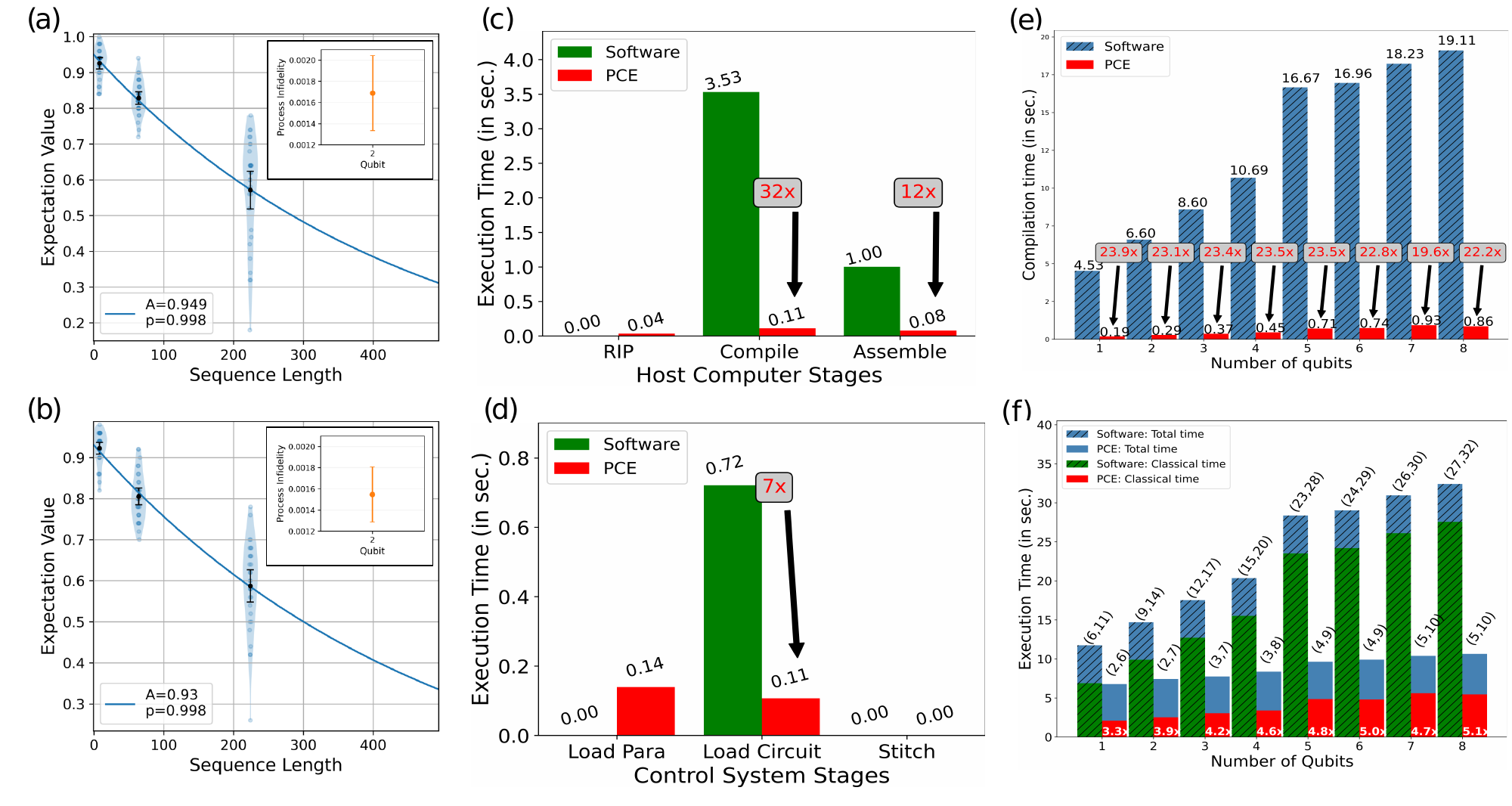}
    \caption{\textbf{Randomized Benchmarking.}
    Single-qubit RB result for \textbf{(a)} software RB and \textbf{(b)} PCE RB. The SPAM (state-preparation and measurement) parameter $A$ and exponential fit parameter $p$ are included in the legends for both plots. The circular data points are the results of individual circuits, and the violin plots depict the distribution of results at each circuit depth. The process infidelity is plotted in the inset in each figure. The two results have a perfect agreement (up to the uncertainty).
    \textbf{(c)} Breakdown of the host computer time profiling for the single-qubit results in (a) and (b). PCE is $27\times$ and $13\times$ faster for the compile and assemble stages, respectively. \textbf{(d)} Breakdown of the control system time profiling for the single-qubit results in (a) and (b). PCE is $7\times$ faster in circuit loading, but has the added cost of loading the parameters. However, there is no runtime overhead for \emph{Stitch}ing. \textbf{(e)} Time profile of (simultaneous) RB for qubit numbers varying from $1$ -- $8$. The plot represents the compilation time on the host computer (\texttt{Compile} and \texttt{Assemble} stages in Section~\ref{subsec:timeprof}).
    \textbf{(f)} Plot represents the total execution time for the software and parameterized execution. The numbers at the bottom shows the total speedup and the numbers in the parentheses show the classical time (all stages described in Section~\ref{subsec:timeprof} minus the \texttt{StartRun} stage) and the total execution time. We observe that the parameterized execution is $\sim 3.3\times$ -- $5.1\times$ faster on classical time, and $\sim 24\times$ faster in compilation.
    }
    \label{fig:rb}
\end{figure*}

Randomized benchmarking (RB), a crucial tool for assessing the performance of quantum gates, is a technique that measures the average error rate over a sequence of randomly chosen Clifford gates (a specific set of quantum gates). A depth $m$ RB circuit consists of $m$ randomly sampled Clifford gates and a single inversion gate at the end that rotates the system back to the starting state. For single-qubit RB, all single-qubit Cliffords can be decomposed into native gates and virtual phases via Equation~\ref{eq:zxzxz}. Thus, a depth $m$ single-qubit RB circuit consists of $2(m+1)$ physical $X_{\pi/2}$ gates and $3(m+1)$ \emph{virtual}-$Z$ gates. This repeated structure makes RB an ideal candidate for PCE since all waveforms for a given circuit depth are structurally equivalent, allowing the PCE design to efficiently compile and execute the circuits. 

In Figure~\ref{fig:rb}(a) and (b), we compare single-qubit RB expectation values measured using the standard software-based procedure, where all circuits are independently sequenced and measured, with the results obtained using PCE. We observe excellent agreement between the two (up to the uncertainty), with process infidelity (i.e., error per Clifford) of $1.7(2) \times 10^{-3}$ and $1.5(1) \times 10^{-3}$ for the software and PCE results, respectively. Additionally, Figures~\ref{fig:rb}(c) and (d) show the profiling results for the single-qubit RB on the host and control systems, respectively. We observe that PCE provides a substantial speedup in various stages of the execution process. On the host computer, PCE achieves a $32\times$ speedup for the compile stage and a $12\times$ speedup for the assemble stage. These improvements are attributed to the reduced compilation overhead, as PCE only needs to compile the unique physical pulse sequences once. On the control system side, we observe a $7\times$ speedup for loading the circuits. However, there is an additional cost associated with loading the parameters onto the FPGA, which partially offsets the speedup in other stages. Importantly, the quantum runtime, represented by the \emph{Stitch} time, remains the same for the same number of shots, even though the PCE circuits are performing \emph{virtual}-$Z$ on the FPGA. 

Figures~\ref{fig:rb}(e) and (f) show the results for software and PCE RB when scaling the number of qubits from 1 to 8, using the experiment parameters described in Appendix~\ref{subsec:paramrb}. As shown in Figure~\ref{fig:rb}(e), the compilation speed up (both \texttt{Compile} and \texttt{Assemble} stage combined) is $\sim 24\times$ compared to software and remains relatively constant with growing number of qubits. Overall, we observe a speedup of $3.3\times$ -- $5.1\times$ in the classical processing time and a $2.5\times$ for the total runtime, as shown in Figure~\ref{fig:rb}(f). This consistent quantum runtime indicates that the hardware-assisted PCE does not introduce any significant overhead, even with the addition of the modified parameter fetch instructions. The speedup in Figure~\ref{fig:rb}(f) is lower compared to Figures~\ref{fig:rb}(c) and (d) due to two factors: the constant data read-back time, which adds approximately $10\%$ of the overall classical time, and the circuit runtime with a passive reset of $500 \mu$s (the time required for the qubit to return to its ground state after measurement), which contributes to $23\%$ of the overall runtime. Despite these factors, we observe that the classical processing time for software and PCE design scales linearly with the number of qubits, it scales significantly more slowly for PCE. This highlights a key advantage of PCE: its ability to maintain execution efficiency as the system scales to larger numbers of qubits.

It is important to note that PCE might not provide significant speedup for all Randomized Benchmarking protocols. For example, in two-qubit RB, decomposing two-qubit Clifford gates into native gates can result in varying numbers of native gate operations. Since any two-qubit unitary can be expressed using one to three native two-qubit gates, resulting in circuits at a given depth not always structurally-equivalent. This variability limits the effectiveness of the PCE design in exploiting the shared structure of these circuits. However, even in such cases, PCE can still offer benefits by optimizing parameter handling and reducing overhead during the load operation, potentially improving overall performance.

\subsection{Cycle Benchmarking}
\begin{figure*}[!t]
    \centering
    \includegraphics[width=0.9\textwidth]{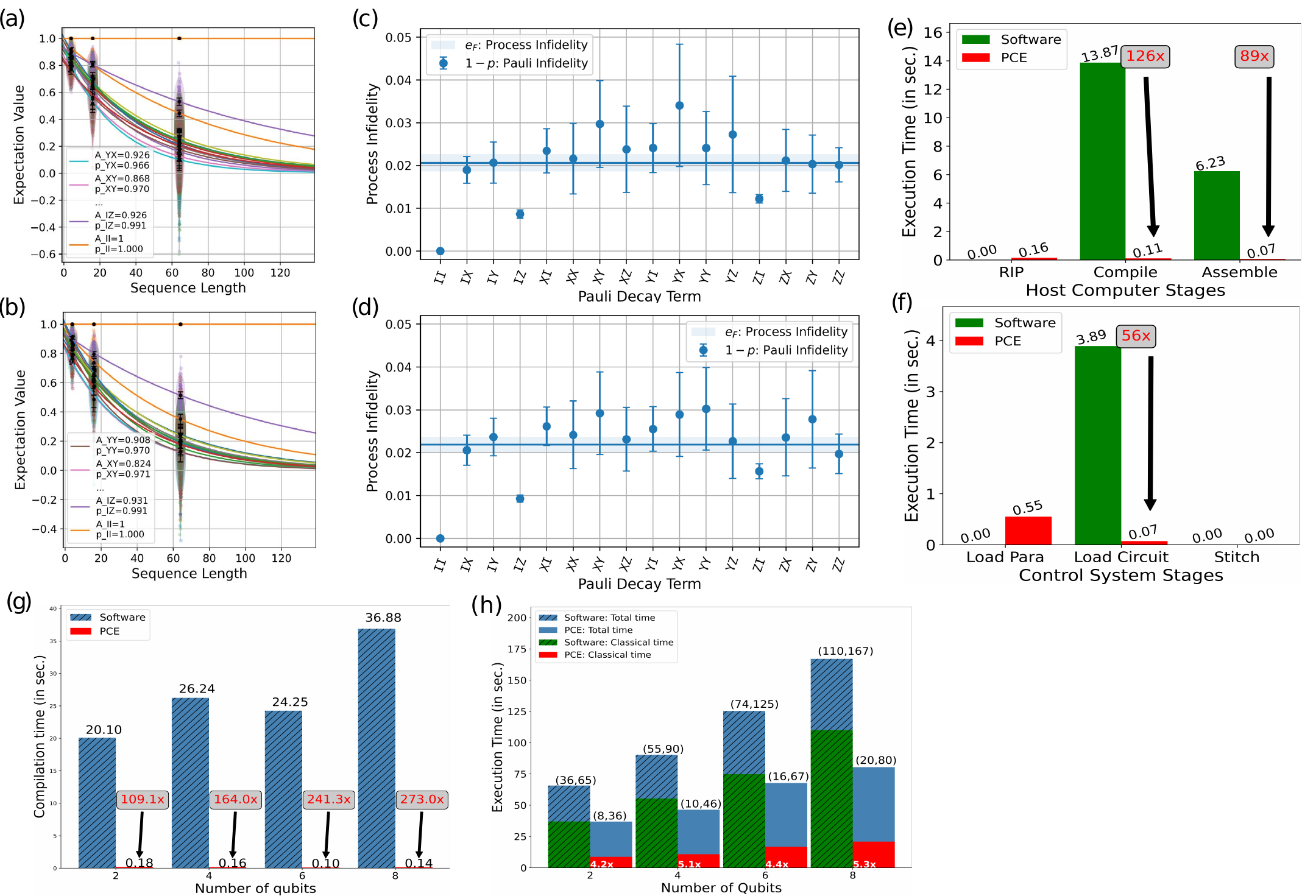}
    \caption{\textbf{Cycle Benchmarking.}
     Exponential decays for \textbf{(a)} software CB and \textbf{(b)} PCE CB performed on a two-qubit CZ gate. Each Pauli decay is fit to an independent exponential function, with $A_P$ and $p_P$ denoting the SPAM (state-preparation and measurement) constant and the exponential fit parameter for each Pauli basis state $P$, respectively. The circular data points are the results of individual circuits, and the violin plots depict the distribution of results for each Pauli decay at each circuit depth. The Pauli infidelities and process infidelity of the gate are plotted in \textbf{(c)} and \textbf{(d)} for the results in (a) and (b), respectively. The process infidelities of both experiments are equal (up to uncertainty). 
     \textbf{(e)} Breakdown of the host computer time profiling for the results in (a) and (b) shows PCE is $252\times$ and $187\times$ faster in the compile and assemble stages, respectively.
     \textbf{(f)} Breakdown of the control system time profiling for the single-qubit results in (a) and (b) shows PCE is $100\times$ faster at the load circuit stage but has the added cost of loading the parameters.  \textbf{(g)} Time profile of CB for qubit numbers varying from $2$ -- $8$. The plot represents the the compilation time on the host computer (\texttt{Compile} and \texttt{Assemble} stages in Section~\ref{subsec:timeprof}). The compilation speed up in the range of $109\times$ -- $273\times$, and increases with larger qubits.
     \textbf{(h)} The plot represents the execution time for the software and PCE execution. The numbers in the parentheses show the classical time (all stages described in Sect.~\ref{subsec:timeprof} minus the \texttt{StartRun} stage) and the total execution time. We observe that the PCE execution is $3.8\times$ -- $4.7\times$ faster.
    }
    \label{fig:cb}
\end{figure*}

Cycle benchmarking (CB) is a technique used in characterizing the performance of quantum gates, particularly multi-qubit gates, by measuring the error rate over a sequence of interleaved cycles of gates. A depth $m$ CB circuit consists of $m$ interleaved cycles of $K$ random Pauli gates followed by an $n$-qubit gate cycle, as well as a single cycle of Paulis at the end of the sequence, which rotates the system back to the initial Pauli eigenstate. This structure repeated over several circuits makes CB suitable for PCE, as all CB circuits of a given depth are structurally equivalent. The single-qubit Pauli gates are decomposed according to Equation~\ref{eq:zxzxz}, while the $n$-qubit gate cycle is decomposed in a manner that depends on the specific gates in the cycle. In this experiment, our interleaved gate is a two-qubit CZ gate, a native gate in our system, which is directly compiled down to its pulse definition. 

We first perform a CB experiment on a two-qubit gate. In Figures~\ref{fig:cb}(a) and (b), we compare exponential decay for CB circuits measured using the standard software-based procedure with the results obtained using PCE, where all circuits are independently sequenced and measured. In Figures~\ref{fig:cb}(c) and (d), we plot the individual Pauli infidelities (which define the different bases in which the system is prepared and measured), as well as the average process infidelity. We find that the software and PCE results are in perfect agreement (up to the uncertainty), with a (dressed) CZ process infidelity of $2.1(1) \times 10^{-2}$ and $2.19(9) \times 10^{-2}$ for the software and PCE results, respectively. Figures~\ref{fig:cb}(e) and (f) show the profiling results for CB on the host and control systems, respectively. On the host computer, PCE achieves a remarkable $126\times$ speedup for the compile stage and a $89\times$ speedup for the assemble stage. These significantly higher speedup values compared to RB are due to CB typically requiring $10\times$ more circuits than RB because of the need to prepare and measure in different bases. On the control system side, we observe a $56\times$ speedup for loading the circuits. Although CB circuits have a parameter loading overhead absent in RB, the combined time for parameter and circuit loading is still at least $5\times$ faster than the corresponding software-based procedure.

Figures~\ref{fig:rb}(e) and (f) show the results for software-based and PCE-based CB when scaling the number of qubits from $2$ to $8$, using the experiment parameters described in Appendix~\ref{subsec:paramcb}. As shown in Figure~\ref{fig:cb}(g), the compilation time speedup increases with the number of qubits, ranging from $109\times$ for $2$ qubits to $273\times$ for $8$ qubits. This improvement is due to the large number of structurally-equivalent circuits in CB, which PCE can efficiently handle. Overall, we observe a $4.2\times$ --- $5.3\times$ speedup in the classical processing time and a $1.8\times$ speedup for the total runtime, as shown in Figure~\ref{fig:cb}(h). As was in the case of RB, the CB performance is affected by the data read-back time, which constitutes $28\%$ of the overall classical time, and the circuit runtime, which accounts for $44\%$ of the total execution time. Moreover, as with RB, the classical runtime improvement with PCE becomes more significant as the number of qubits scale from 2 to 8. This demonstrates the remarkable scalability and efficiency of the hardware-assisted PCE design, particularly for larger quantum systems.

\subsection{Gate Set Tomography}

\begin{table*}[!t]
    \centering
    \caption{\textbf{Comparison of GST Metrics with the standard GST run on the software and with PCE.}
    }
        \begin{tabular}{l || c | c}
            & Standard GST & GST with PCE \\
            \hline
            \hline
            Entanglement Inf. ($X_{\pi/2}$ on Q2) & $0.00446 \pm 0.000074$ & $0.003901 \pm 0.00007$ \\
            Entanglement Inf. ($Y_{\pi/2}$ on Q2) & $0.004187 \pm 0.000076$ & $0.003060 \pm 0.000073$ \\
            Entanglement Inf. ($X_{\pi/2}$ on Q3) & $0.002952 \pm 0.000074$ & $0.003238 \pm 0.000066$ \\
            Entanglement Inf. ($Y_{\pi/2}$ on Q3) & $0.003428 \pm 0.000081$ & $0.003990 \pm 0.000083$ \\
            Entanglement Inf. (CZ) & $0.021487 \pm 0.000284$ & $0.020235 \pm 0.000192$ \\
            Model Violation ($N_\sigma$ @ $L=128$) & 2670.93 & 411.152 \\
        \end{tabular}
\label{tab:gst}
\end{table*}

\begin{figure*}[!t]
    \centering
    \includegraphics[width=0.75\textwidth]{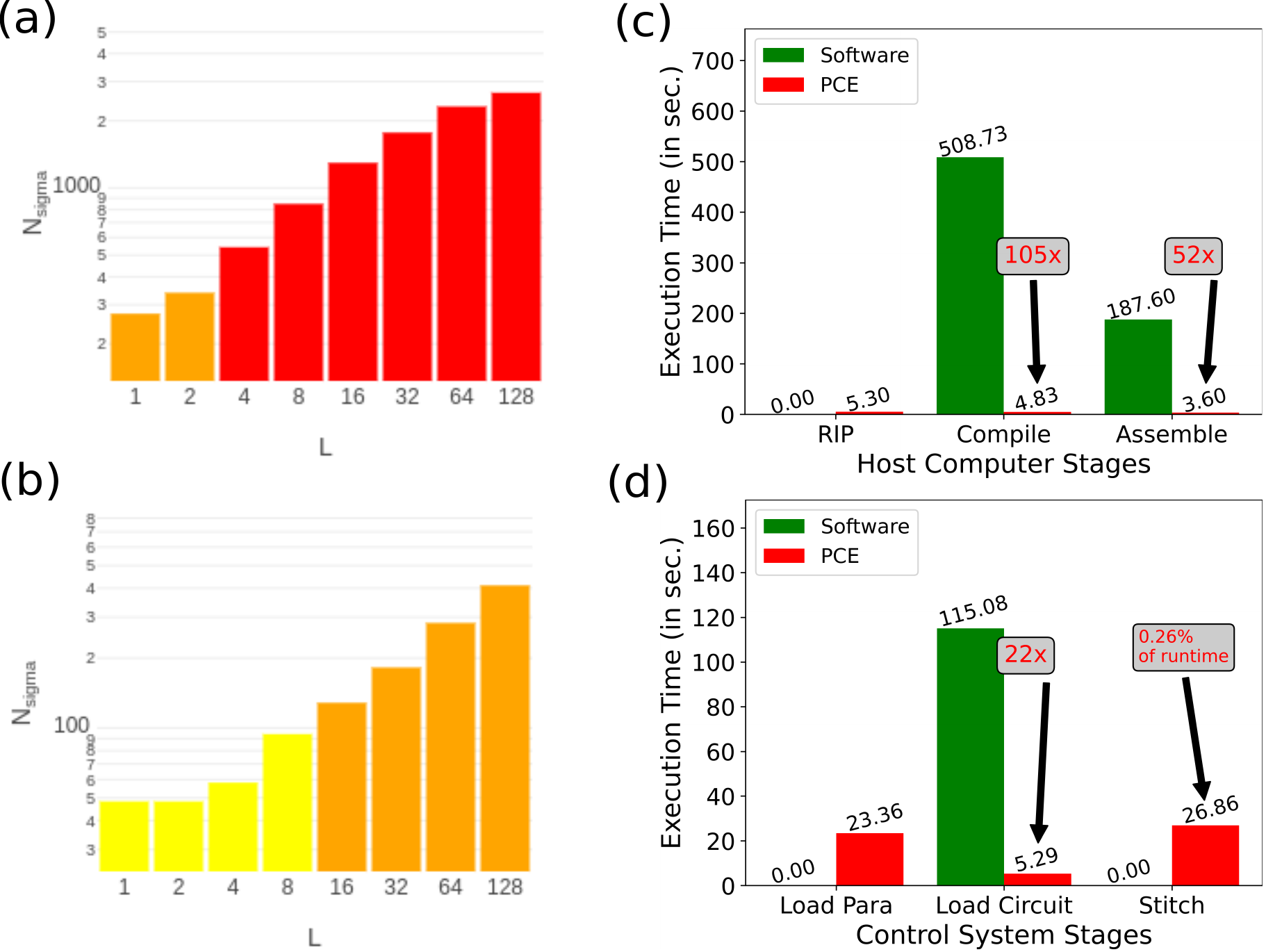}
    \caption{\textbf{Gate Set Tomography.}
    Comparison of model violation of GST with \textbf{(a)} Software and \textbf{(b)} PCE. The PCE results have less model violations compared to software 
    Time profiling for GST with and without PCE  on the \textbf{(c)} host side and \textbf{(d)} control side. PCE is $105\times$ and $52\times$ faster for the compile and assemble stages, respectively, and $22\times$ faster at the load circuit stage.
    }
    \label{fig:gst}
\end{figure*}

Tomographic reconstruction methods, such as QPT and GST, scale exponentially in the number of qubits and are thus computationally expensive form of gate characterization. While PCE cannot change the inherent poor scaling of tomography, it can improve the runtime performance by exploiting the high degree of structural redundancy in circuits, especially in long-sequence GST. To demonstrate this, we perform two-qubit GST in sooftware and with PCE (experiment parameters are described in Appendix~\ref{subsec:paramgst}) and summarize key performance metrics in Table \ref{tab:gst}. We observe that the estimated entanglement infidelities for all of the gates in the gate set (except $Y_{\pi/2}$ on Q2) are equal up to $10^{-3}$ for the single-qubit gates and $10^{-2}$ for two-qubit CZ gate, respectively. Additionally, as shown in Figure~\ref{fig:gst}(a) and (b), we observe that the model violation ($N_\sigma$) is significantly lower for PCE (411) compare to the standard case (2672), suggesting that the GST model for the PCE data is more trustworthy. This could be, in part, due to the fact that PCE reduces the total execution time, and thus would be less susceptible to any kind of parameter drifts in QPU during the execution of the circuits.

In Figure~\ref{fig:gst}(c) and (d) show the profiling results for GST on the host computer and control systems, respectively. On the host computer, PCE achieves a $105\times$ speedup for the compile stage and a $52\times$ speedup for the assemble stage. For GST, we employ out-of-order structural matching, meaning that the circuits are not executed in their original sequence. This is necessary because, unlike other quantum characterization, verification, and validation (QCVV) protocols where adjacent circuits are similar, GST circuits do not have structural equivalency between adjacent circuits. However, as shown by the lower model violation count for PCE in Figure~\ref{fig:gst}(b), this circuit reordering does not degrade, but improves the confidence of the GST results. On the control system side, PCE achieves a $22\times$ speedup for loading the circuits. This speedup is due to the reduced number of circuits loaded with PCE (251) compared to the standard procedure (19488), as shown in Table~\ref{tab:summary}. The GST circuits consists of $4,091,346$ parameters,  which requires a significant time to load, as shown in Figure~\ref{fig:gst}(d). However, the combined time for parameter and circuit loading is still at least $4\times$ faster than the corresponding software-based procedure. Additionally, the \emph{Stitch} time is $26.86$ seconds with PCE for $1000$ shots, resulting from stitching of $4$ billion parameters ($4,091,346$ parameters over $1000$ shots). However, this overhead is about $0.26\%$ of the total runtime. Overall, PCE provides a $6.86\times$ speedup in the classical processing time as shown in Table~\ref{tab:summary}, demonstrating its effectiveness in accelerating even complex tomographic reconstruction methods. 

\subsection{Summary}
In this section, we validated the effectiveness of the hardware-assisted PCE design across a range of quantum algorithm benchmarking tasks. We measure the advantage of PCE using compilation speedup given in Eq.~\ref{eq:compsu}, percentage of classical time reduction given in Eq.~\ref{eq:ctr}, and classical speedup given in Eq.~\ref{eq:csu}.

\begin{table*}
\begin{equation}
    \label{eq:compsu}
    \texttt{Compilation Speedup} = \frac{(\texttt{Software Compile Time} + \texttt{Software Assemble Time})}{(\texttt{PCE Compile Time} + \texttt{PCE Assemble Time})}
\end{equation}

\begin{equation}
\label{eq:ctr}
\begin{split}
    \texttt{Classical Time Reduction (\%)} = \frac{(\texttt{Software Classical Time}-\texttt{PCE Classical Time})}{\texttt{Software Classical Time}} * 100 \\
    \texttt{Classical Time} = \texttt{Total execution time} - \texttt{Quantum circuit runtime}
\end{split}
\end{equation}

\begin{equation}
\label{eq:csu}
    \texttt{Classical Speedup} = \frac{\texttt{Software Classical Time}}{\texttt{PCE Classical Time}}
\end{equation}
\end{table*}

The results, summarized in Table~\ref{tab:summary}, demonstrate significant performance improvements compared to standard software-based procedures. The table highlights that these algorithms exhibit a large degree of structural equivalency, either in-order or out-of-order (as in the case of GST), as evidenced by the substantial difference between the total number of circuits and the number of \emph{unique circuits}. For Randomized Compiling (RC), PCE achieved compilation speedups of $17.48\times$ and $995.05\times$ for RC20 and FRC resulting in classical time reductions of $64.39\%$ and $77.22\%$, respectively. Randomized Benchmarking (RB) showed a $22.34\times$ compilation speedup and a $77.87\%$ reduction in classical processing time. Cycle Benchmarking (CB) exhibited even more substantial gains, with a $185.34\times$ compilation speedup and an $77.22\%$ reduction in classical time. Gate Set Tomography (GST), a particularly challenging protocol due to its high circuit complexity, benefited significantly from PCE, achieving an $82.65\times$ compilation speedup and an $85.42\%$ reduction in classical processing time. The compilation speedup, calculated as the time taken for compiling and assembling the circuits using the \texttt{QubiC} compiler, closely approaches the theoretical limit determined by the ratio of the total circuits to the \emph{unique circuits}. The efficient handling of parameter stitching, involving hundreds of thousands to millions of parameters, relies on the hardware-accelerated \emph{Stitch} module. Across all protocols, PCE consistently reduced the classical processing time (calculated as the total time minus the quantum runtime, as described in Section 3), with overall speedups ranging from $2.81\times$ to $6.86\times$. While hardware-assited PCE reduces the compilation and circuit loading time with minimal overhead in runtime and parameter loading, the overall quantum algorithm execution time is also influenced by other classical operations within the control system, such as data readout and data management between the ARM and FPGA which influence the total classical time. Additionally, practical quantum systems require classical system support for hundreds of qubits which require advancement in quantum algorithms~\cite{PRXQuantum.5.030334} and control systems~\cite{huangQCE2024}. However, from the Table~\ref{tab:stitchutil}, we can infer that the \emph{Stitch} module is extensible with the control system architecture due to the low resource utilization and a preliminary scalability study on \emph{RIP} (in Appendix~\ref{sec:scale}) shows compilation gain for up to $20$ qubits. These results highlight the ability of hardware-assisted PCE to accelerate the execution of quantum algorithms by minimizing compilation overhead and efficiently managing circuit parameters, particularly in scenarios involving large numbers of structurally-equivalent circuits.

\begin{table*}[t]
    \centering
    \caption{\textbf{Impact of PCE with classical time speedup for various protocols.} The data shows the total number of circuits, the percentage of classical time, the structurally-equivalent circuits identified by \emph{RIP}, classical time reduction, and the speedup of PCE for each experiment based on the parameters described in Appendix~\ref{sec:expparam}. The structural equivalency, identified by \emph{RIP}, shows that the circuits in the protocols have a significant similarities with only $<5\%$ of \emph{unique circuits}. The classical time is the percentage of total time excluding the quantum run time with a passive reset of $500 ns$. The classical time reduction shows the impact, reducing the overall classical time by $64\%$ to $85\%$ depending on the protocol. Additionally, the overall speedup considering the run time with the classical time varies between $2.81\times$ to $6.86\times$ showing a substantial reduction in quantum circuit execution.}
    \begin{tabularx}{1.0\textwidth} { 
      >{\raggedright\arraybackslash}X 
      || >{\raggedleft\arraybackslash}X 
      | >{\raggedleft\arraybackslash}X 
      | >{\raggedleft\arraybackslash}X 
      | >{\raggedleft\arraybackslash}X 
      | >{\raggedleft\arraybackslash}X 
      | >{\raggedleft\arraybackslash}X }
      \textbf{Protocol}&\textbf{Number of Circuits}&\textbf{Number of \emph{Unique Circuits}}&\textbf{Parameters}&\textbf{Compilation Speedup ($\times$)}&\textbf{Classical Time Reduction(\%)}&\textbf{Classical Speedup ($\times$)}\\
    \hline
    \hline
    RC20 & 1,540 & 77 & 1,180,200 &17.48 & 64.39 & 2.81\\
    FRC & 77,000 & 77 & 59,010,000& 995.05 & 77.22 & 4.39\\
    CB & 3,240 & 12 & 462,840 & 185.34 & 77.23 & 4.39\\
    RB &  736  & 32 & 638,496&22.34 & 77.87 & 4.52\\
    GST & 19,488 & 251 & 4,091,346&82.65 & 85.42 & 6.86\\
    \end{tabularx}
    \label{tab:summary}
\end{table*}
\section{Conclusions}\label{sec:conclusions}

In this work, we developed hardware-assisted PCE, a hardware-software co-design for efficient execution of parameterized quantum circuits integrated with open-source FPGA-based control hardware, \texttt{QubiC}. The design introduces \emph{Read-Identify-Peel (RIP)} and \emph{Stitch} with \emph{Deft Scheduler}, to identify structurally-equivalent circuits, efficiently manage and store circuit parameters on the FPGA, and stitch the parameters back to generate analog waveforms for quantum circuit execution. The design is scalable and uses minimal FPGA resources as shown in utilization analysis. The results demonstrate significant time savings for various quantum characterization, verification, and validation (QCVV) protocols, as highlighted in Table~\ref{tab:summary}. Depending on the protocol, PCE reduced compilation time by $17.42\times$ to $995.05\times$ (close to the theoretical limit), classical processing time by $64.39\%$ to $85.42\%$, and achieved classical speedups of $2.81\times$ to $6.86\times$. These speedups can substantially accelerate quantum algorithm development without violating the integrity of device characterization. 

More importantly, the current design is algorithm-agnostic and wraps the entire functionality in a single API call. While demonstrated for typical QCVV workflows, hardware-assisted PCE's parameterization capabilities can be extended to many classes of circuits. For example, applying PCE to variational quantum algorithms~\cite{cerezo2021variational} could significantly reduce the overhead of parameter optimization loops. Similarly, adapting PCE to quantum gate calibration, which involves iterative sweeps of pulse parameters, could dramatically improve the efficiency of device bring-up procedures. We believe that PCE exemplifies the potential of co-designing classical control hardware to minimize overhead in quantum workflows. Future work will explore further opportunities to optimize quantum computations through co-design for larger number of qubits and different classes of quantum algorithms.
\paragraph*{\textbf{\textup{Acknowledgments}}}\label{sec:acknowledgements}
The majority of this work was supported by the Laboratory Directed Research and Development Program of Lawrence Berkeley National Laboratory under U.S. Department of Energy Contract No.~DE-AC02-05CH11231. Y.X., I.S., and G.H.~acknowledge financial support for the primary development of the \texttt{QubiC} hardware from the U.S.~Department of Energy, Office of Science, Office of Advanced Scientific Computing Research Quantum Testbed Program under Contract No.~DE-AC02-05CH11231 and the Quantum Testbed Pathfinder Program.

\paragraph*{\textbf{\textup{Author Contributions}}}\label{sec:author_contributions}
A.D.R. designed the RIP and Stitch method and performed the experiments. A.D.R. and A.H.~performed the data analysis. A.D.R, N.F., Y.X., and G.H.~developed the control system \texttt{QubiC} used in this work. J.H.~assisted in the design and analysis of the GST data. A.D.R.~and A.H.~wrote the manuscript with input from all coauthors. G.H., I.S., K.K., and K.N.~supervised the work.

\paragraph*{\textbf{\textup{Competing Interests}}}\label{sec:competing_interests}
The hardware-assisted parameterized circuit execution presented in this work is protected under the U.S.~patent application no.~PCT/US25/30612, filed by Lawrence Berkeley National Laboratory on behalf of the following inventors: A.D.R., N.F., A.H., and K.N.

\paragraph*{\textbf{\textup{Data Availability}}}\label{sec:data_availability}
All data are available from the corresponding author upon reasonable request.

\bibliography{bibliography}

\clearpage
\appendix
\setcounter{table}{0}
\renewcommand{\thetable}{A\arabic{table}}

\setcounter{figure}{0}
\renewcommand{\thefigure}{A\arabic{figure}}
\clearpage
\section{Experiment Parameters}
\label{sec:expparam}
Below are the experiment parameters used for profiling in Sections~\ref{sec:time} and~\ref{sec:validation}. Each circuit has a delay of $500\mu s$ per shot. The \texttt{depths} represent the depth of the circuit, the \texttt{widths} represent the number of physical qubits, total circuits represent sum of all the circuits in the entire experiment covering all \texttt{depths}, \texttt{widths}, \texttt{randomizations} (for RC 20 and FRC), and \texttt{shots} represent the number of repetitive measurements per circuit. 

\subsection{Randomized Compiling - RC20}
\label{subsec:paramrc20}
    \begin{itemize}
        \item \textbf{depths} = [1, 10, 20, 30, 40, 50, 60, 70, 80, 90, 100]
        \item \textbf{widths} = [(0,1), (0,1,2), (0,1,2,3), (0,1,2,3,4), (0,1,2,3,4,5),  (0,1,2,3,4,5,6), (0,1,2,3,4,5,6,7)]
        \item \textbf{randomization per depth and width} = 20
        \item \textbf{Total number of circuits} = randomization $\times$ depth $\times$ width $20\times11\times7 = 1540$
        \item \textbf{shots per depth and width} = 50 per randomization = $50\times20 = 1000$
    \end{itemize}
\subsection{ RC Fully Randomly Compiled - RC FRC}
\label{subsec:paramfrc}
\begin{itemize}
    \item \textbf{depths} = [1, 10, 20, 30, 40, 50, 60, 70, 80, 90, 100]
    \item \textbf{widths} = [(0,1), (0,1,2), (0,1,2,3), (0,1,2,3,4), (0,1,2,3,4,5), (0,1,2,3,4,5,6), (0,1,2,3,4,5,6,7)]
    \item \textbf{randomization per depth and width} = 1000
    \item \textbf{Total number of circuits} =  randomization $\times$ depth $\times$ width = $1000\times11\times7 = 7700$
    \item \textbf{shots per depth and width} = 1 per randomization = $1\times1000 = 1000$
\end{itemize}

\subsection{Cycle Benchmarking - CB}
\label{subsec:paramcb}
\begin{itemize}
    \item \textbf{widths} = [(0,1), (0,1,2,3), (0,1,2,3,4,5), (0,1,2,3,4,5,6,7)]
    \item \textbf{randomization} = [[4, 16, 64], [4, 8, 32], [2, 4, 8], [2, 4, 8]]
    \item \textbf{number of circuits per width} = [540, 660, 960, 1080]
    \item \textbf{Total number of circuits} = $540+640+960+1080 = 3240$
    \item \textbf{shots per circuit} = 100
\end{itemize}

\subsection{Randomized Benchmarking - RB}
\label{subsec:paramrb}
\begin{itemize}
    \item \textbf{widths} = [(0),(0,1), (0,1,2), (0,1,2,3), (0,1,2,3,4), (0,1,2,3,4,5), (0,1,2,3,4,5,6), (0,1,2,3,4,5,6,7)]
    \item \textbf{randomization} = [(16,128,384), (16,96,384), (16,64,256), (16,64,192), (8,64,192), (8,32,160), (4,32,160), (4,32,128)]
    \item \textbf{number of circuits per width} = (randomization $\times$ number of circuit per randomization) + read circuits 
    \item[] = ${3\times 30 + 2 = 90 + 2 = 92}$
    \item \textbf{Total number of circuits} = circuit per width $\times$ width = $92\times8 = 736$
    \item \textbf{shots per circuit} = 100
\end{itemize}
\subsection{Gate Set Tomography - GST}
\label{subsec:paramgst}
\begin{itemize}
    \item \textbf{depths} = [2, 4, 8, 16, 32, 64, 128]
    \item \textbf{widths} = [(0,1)]
    \item \textbf{Total number of circuits} = 19488
    \item \textbf{shots per circuit} = 1000
\end{itemize}

\section{Code Snippets}
Below are the Python code snippets for the application and \emph{RIP} process.
\label{sec:code}
\subsection{Stitch module instructions}
\label{subsec:stitchcode}
\begin{lstlisting}[label={lst:modc},caption={\texttt{QubiC} specific instruction added for each parameterized gate instructing to fetch the parameter from Stitch module on FPGA.}, style={abstyle}]
#create a variable
{'name': 'declare', 'var': varname, 'dtype': 'phase', 'scope': [i]}
#bind parameter to the qubit frequency
{'name': 'bind_phase', 'freq': f'{i}.freq', 'var': varname},
#set intial value to 0
{'name': 'set_var', 'value':0, 'var': varname}] 
# Get parameter from stitch module
{'name': 'alu_fproc','func_id':10, 'op':'add', 'lhs':varname,'out':varname},
# processing delay for 'Stitch' logic overlapped with gate execution
{'name': 'delay', 't': 4.e-9}
\end{lstlisting}

\subsection{RIP module signature}
\label{subsec:ripcode}
\begin{lstlisting}[label={lst:pce},caption={method to perform read-identify-peel process from user application.}, style={abstyle}]
def pce(circuits, qubits, chanmap):
    """
    Method to detect structurally similar circuits from a batch of circuits
    and extract the phases
    Args:
        circuits (List): List of circuits transpiled to QubiC format
        qubits: Qubits index as integer
        chanmap: Physical mapping of qubits
    Returns:
        Dict: Parameter dictionary for Stitch
        List: Unique circuits from structurally-equivalent circuits modified
        for runtime phase matching
    """
\end{lstlisting}

\subsection{Build and run code}
\label{subsec:runcode}
\begin{lstlisting}[label={lst:buildrun},caption={Modification of build and run method from \texttt{QubiC} API to support parameterization.}, style={abstyle}]
def build_and_run_paracircuits(self, 
    program_list: List, #list of QubiC circuits
    parameters: Dict, #parameter dictionary
    server_profile: int = 0, #support for profiling
    ... #rest of the arguments from QubiC API build_and_run_circuits
    ...) -> dict:
\end{lstlisting}

\subsection{Profile and run application}
\label{subsec:profcode}
\begin{lstlisting}[label={lst:profile},caption={A demo application showing use of profiling infrastructure in the application layer}, style={abstyle}]
import profiling as tm #import porfiling module
tm.setClientProf() #Enable client-side profiling
profile_level = 1 #Enable server-side profiling
#Perform quantum experiment
build_and_run_paracircuit(circuit, parameters, server_profile=profile_level) #Run with profiling
tm.logjson() #Log profile dictionary in JSON format
\end{lstlisting}

\subsection{Circuit Modification}
\label{subsec:circmod}
The below example shows the circuits compiled using Equation~\ref{eq:zxzxz} into \texttt{QubiC} native set instructions  The Listing~\ref{lst:origcirc} shows two structurally-similar circuits but vary only by the \emph{virtual}-$Z$ phase values. The Listing~\ref{lst:modicirc} shows the modification by \emph{RIP} to create \emph{unique} circuit that serves as a template for both the circuits in Listing~\ref{lst:origcirc}.

\begin{minipage}[t]{1.0\linewidth}
\begin{lstlisting}[style={abstyle}, caption={An example structurally-similar circuits compiled using Equation~\ref{eq:zxzxz}.}, numbers=none, label={lst:origcirc}]
[[{'name':'delay','t':0.0005}, #Circuit 1
{'name':'virtual_z','qubit':['Q0'],'phase':-0.1633},
{'name':'virtual_z','qubit':['Q1'],'phase':-1.3352},
{'name':'CNOT','qubit':['Q0','Q1']},
{'name':'virtual_z','qubit':['Q0'],'phase':-4.5092},
{'name':'virtual_z','qubit':['Q1'],'phase':1.1853},
{'name':'X90','qubit':['Q0']},
{'name':'X90','qubit':['Q1']},
{'name':'read','qubit':['Q0']},
{'name':'read','qubit':['Q1']}],
[{'name':'delay','t':0.0005}, #Circuit 2
{'name':'virtual_z','qubit':['Q0'],'phase':-2.1092},
{'name':'virtual_z','qubit':['Q1'],'phase':-1.9562},
{'name':'CNOT','qubit':['Q0','Q1']},
{'name':'virtual_z','qubit':['Q0'],'phase':0.2793},
{'name':'virtual_z','qubit':['Q1'],'phase':0.9643},
{'name':'X90','qubit':['Q0']},
{'name':'X90','qubit':['Q1']},
{'name':'read','qubit':['Q0']},
{'name':'read','qubit':['Q1']}]]
\end{lstlisting}
\end{minipage}%
\hspace{3mm}
\begin{minipage}[t]{1.\linewidth}
\begin{lstlisting}[style={abstyle}, caption={\emph{RIP} modified \emph{unique} circuit with instructions to fetch parameters from \emph{Stitch} module.}, numbers=none, label={lst:modicirc}]
[{'name':'declare','var':'pQ0','dtype':'phase','scope':['Q0']},
{'name':'bind_phase','freq':'Q0.freq','var':'pQ0'},
{'name':'set_var','value':0,'var':'pQ0'},
{'name':'declare','var':'pQ1','dtype':'phase','scope':['Q1']},
{'name':'bind_phase','freq':'Q1.freq','var':'pQ1'},
{'name':'set_var','value':0,'var':'pQ1'},
{'name':'delay','t':0.0005},
{'name':'alu_fproc','func_id':10,'op':'add','lhs':'pQ0','out':'pQ0'},
{'name':'delay','t':4e-09},
{'name':'alu_fproc','func_id':10,'op':'add','lhs':'pQ1','out':'pQ1'},
{'name':'delay','t':4e-09},
{'name':'CNOT','qubit':['Q0','Q1']},
{'name':'alu_fproc','func_id':10,'op':'add','lhs':'pQ0','out':'pQ0'},
{'name':'delay','t':4e-09},
{'name':'alu_fproc','func_id':10,'op':'add','lhs':'pQ1','out':'pQ1'},
{'name':'delay','t':4e-09},
{'name':'X90','qubit':['Q0']},
{'name':'X90','qubit':['Q1']},
{'name':'read','qubit':['Q0']},
{'name':'read','qubit':['Q1']}]
\end{lstlisting}
\end{minipage}

\begin{figure*}
\begin{subfigure}{0.48\textwidth}
    \centering
    \includegraphics[width=1.0\linewidth]{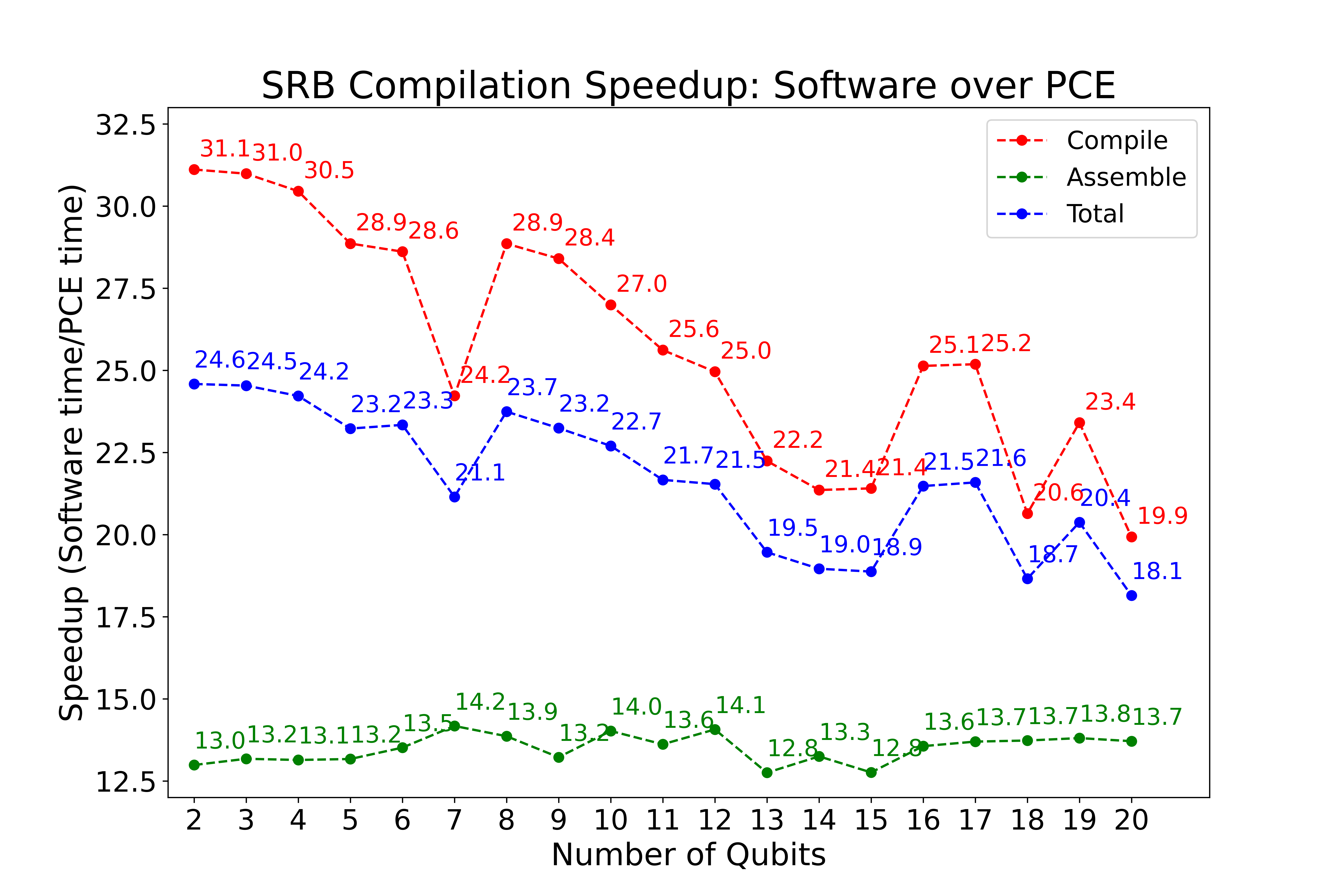}
    \caption{Compilation speedup comparison of Randomized Bechmarking for qubits 2 -- 20 shows hardware-assisted PCE with an assemble time speedup of $12.8\times$ to $14.2\times$, compilation speedup between $19.9\times$ to $31.1\times$, and overall speedup of $18.1\times$ to $24.6\times$. }
    \label{fig:rbcomp}
\end{subfigure}
\hspace{5mm}
\begin{subfigure}{0.48\textwidth}
    \centering
    \includegraphics[width=1.0\linewidth]{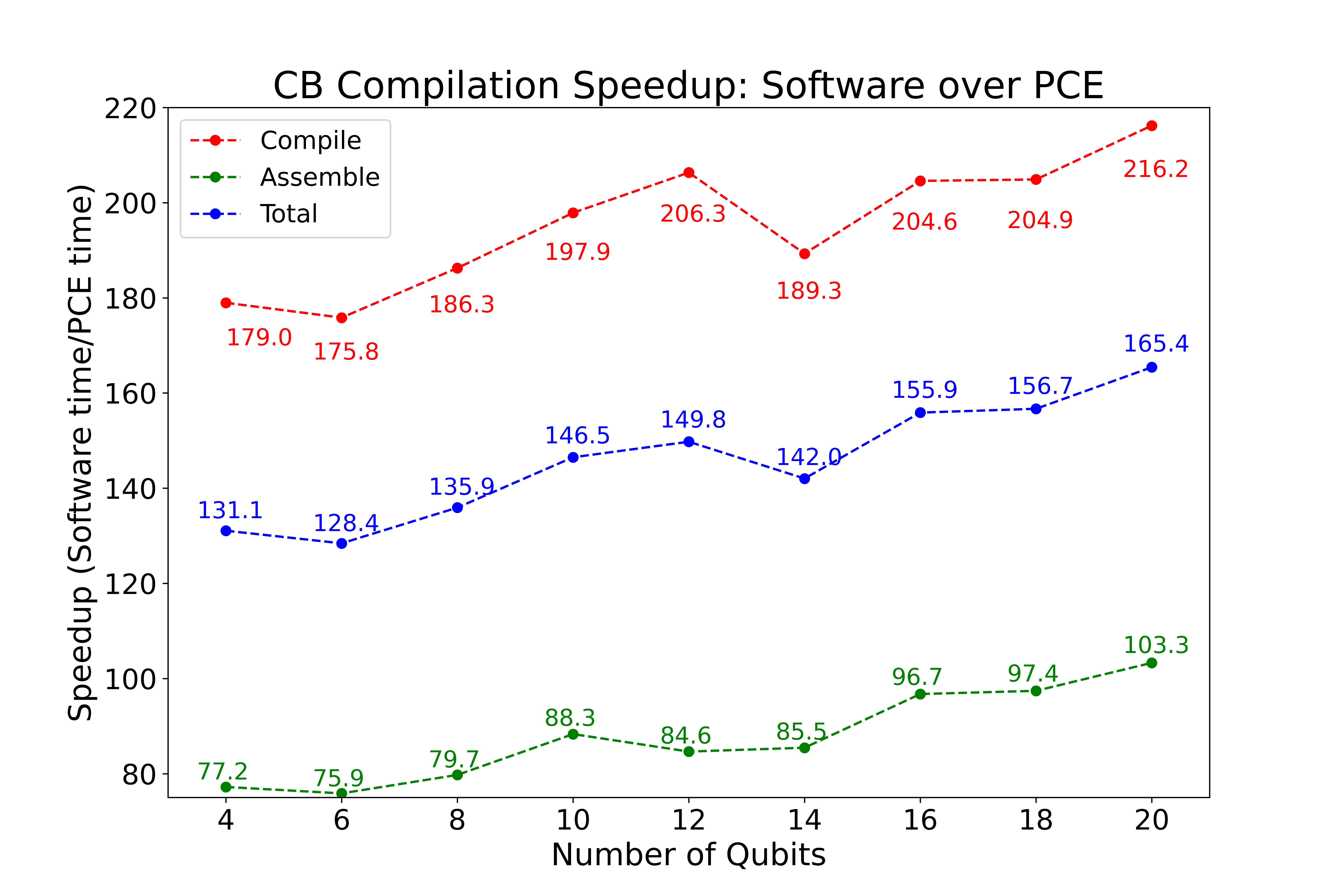}
    \caption{Compilation speedup comparison of Cycle Bechmarking for qubits 4 -- 20 shows hardware-assisted PCE with an assemble time speedup of $77.2\times$ to $103.3\times$, compilation speedup between $179.0\times$ to $216.2\times$, and overall speedup of $131.1\times$ to $165.4\times$. }
    \label{fig:cbcomp}
\end{subfigure}
\caption{Scalability study comparing the compilation time speedup using hardware-assisted PCE and software for Randomized Benchmarking and Cycle Benchmarking protocol for up to $20$ qubits.}
\label{fig:scale}
\end{figure*}

\section{Scalability study of \emph{RIP} module}
\label{sec:scale}

This section describes the compilation speedup of \emph{RIP} compared with the software for up to $20$ qubits for Streamlined Randomized Benchmarking (SRB) and Cycle Benchamrking (CB) protocols. The randomization parameters and the number of circuits for each qubit for SRB and CB are as follows:

\begin{enumerate}
    \item  \textbf{Streamlined Randomized Benchmarking - SRB}
    \label{subsec:sclaerb}
        \begin{itemize}
            \item \textbf{randomization} = [[512, 256, 128, 64, 16], [492, 246, 124, 62, 16], [472, 236, 120, 60, 16], [452, 226, 116, 58, 14], [432, 216, 112, 58, 14], [410, 206, 106, 56, 14], [390, 196, 102, 54, 14], [370, 186, 98, 52, 14], [350, 176, 94, 50, 12], [330, 166, 90, 48, 12], [310, 154, 86, 48, 12], [290, 144, 82, 46, 12], [270, 134, 78, 44, 10], [250, 124, 74, 42, 10], [230, 114, 70, 40, 10], [208, 104, 64, 38, 10], [188, 94, 60, 38, 10], [168, 84, 56, 36, 8], [148, 74, 52, 34, 8], [128, 64, 48, 32, 8]]
            \item \textbf{number of circuits per width} = (randomization $\times$ number of circuit per randomization) + read circuits 
            \item[] = ${5\times 30 + 2 = 150 + 2 = 152}$
        \end{itemize}
        
    \item \textbf{Cycle Benchmarking - CB}
    \label{subsec:scalecb}
        \begin{itemize}
            \item \textbf{randomization} = [[4, 8, 16, 32, 64], [4, 8, 14, 30, 58], [4, 8, 14, 26, 54], [4, 6, 12, 24, 48], [4, 6, 12, 22, 42], [2, 6, 10, 18, 38], [2, 6, 10, 16, 32], [2, 4, 8, 14, 26], [2, 4, 8, 10, 22], [2, 4, 6, 8, 16]]
            \item \textbf{number of circuits per width} = [900, 1400, 1500, 1700, 1900, 1900, 1900, 2000, 2000, 2000]
        \end{itemize}
\end{enumerate}

The speedup is calculated by measuring the compile and assemble stages, as described in Section~\ref{subsec:timeprof} in \emph{Host Computer Software Layer}. As shown in Figure~\ref{fig:scale}, the speedup of \emph{RIP} varies between SRB and CB. For SRB, the assemble time remains relatively constant with average speedup of $13.45$. However, the compilation time varies between $19.9\times$ to $31.1\times$ depending on the number of qubits due to the complexity of the RB circuits. For CB, the compilation speed increases with the number of qubits withan assemble time speedup of $77.2\times$ to $103.3\times$, compilation speedup between $179.0\times$ to $216.2\times$, and overall speedup of $131.1\times$ to $165.4\times$.

\end{document}